\documentclass{agujournal2019}
\usepackage{apacite}
\usepackage{url} 
\usepackage{lineno}
\usepackage[finalnew]{trackchanges} 
\usepackage{soul}
\usepackage{multirow}
\usepackage{enumitem}
\usepackage{rotating}
\usepackage{color}
\usepackage{comment}
\usepackage{adjustbox}
\usepackage{todonotes}
\definecolor{nasa red}{rgb}{0.988235294117647, 0.23921568627451, 0.129411764705882}
\definecolor{nasa blue}{rgb}{0.0431372549019608, 0.23921568627451, 0.568627450980392}
\definecolor{nasa green}{rgb}{0, 0.505882352941176, 0.282352941176471}
\definecolor{nasa orange}{rgb}{0.95294117647, 0.71372549019, 0.12156862745} 
\definecolor{Pantone blue 1}{rgb}{0.21176471, 0.5254902 , 0.62745098} 
\definecolor{Pantone blue 2}{rgb}{0.09803922, 0.31764706, 0.56470588} 
\definecolor{Pantone gold}{rgb}{0.45882353, 0.47843137, 0.30588235}

\draftfalse

\journalname{Space Weather}

\begin{document}

%
%


\title{Toward a next generation particle precipitation model: Mesoscale prediction through machine learning (a case study and framework for progress)}

%
%




\authors{Ryan M. McGranaghan\affil{1}, Jack Ziegler\affil{1}, T\'eo Bloch\affil{2}, Spencer Hatch\affil{3}, Enrico Camporeale\affil{4,5}, Kristina Lynch\affil{6}, Mathew Owens\affil{2}, Jesper Gjerloev \affil{3,7}, Binzheng Zhang\affil{8}, Susan Skone\affil{9}}


\affiliation{1}{Atmosphere and Space Technology Research Associates (ASTRA), Louisville, CO, USA}
\affiliation{2}{University of Reading, Reading, England}
\affiliation{3}{Birkeland Center, University of Bergen, Bergen, Norway}
\affiliation{4}{CIRES, University of Colorado, Boulder, CO, USA}
\affiliation{5}{NOAA Space Weather Prediction Center, Boulder, CO, USA}

\affiliation{6}{Department of Physics and Astronomy, Dartmouth College, Hanover, NH, USA}
\affiliation{7}{Applied Physics Laboratory, The Johns Hopkins University, Laurel, MD, USA}
\affiliation{8}{Department of Earth Sciences, University of Hong Kong, Hong Kong}
\affiliation{9}{University of Calgary, Calgary, Alberta, Canada}





\correspondingauthor{Ryan M. McGranaghan}{ryan.mcgranaghan@gmail.com}




\begin{keypoints}
\item We utilize a data-driven organization of input parameters to produce a new total electron energy flux nowcast model (\textbf{see Figure \ref{model during training figure}})
\item Extended input features and higher expressive power provided by machine learning approach yields more capable mesoscale and peak flux specification (\textbf{see Figure \ref{january 25, 2000 event figure}}) and an overall reduction in specification errors compared with the state-of-the-art (\textbf{see Table \ref{metrics table}})
\item A framework for evaluation of machine learning (and any model) in geospace is suggested, building on momentum in the community (\textbf{see Section \ref{interrogation section}}) 
\end{keypoints}

%
%


\begin{abstract}
We advance the modeling capability of electron particle precipitation from the magnetosphere to the ionosphere through a new database and use of machine learning tools to gain utility from those data. We have compiled, curated, analyzed, and made available a new and more capable database of particle precipitation data that includes 51 satellite years of Defense Meteorological Satellite Program (DMSP) observations temporally aligned with solar wind and geomagnetic activity data. The new total electron energy flux particle precipitation nowcast model, a neural network called PrecipNet, takes advantage of increased expressive power afforded by machine learning approaches to appropriately utilize diverse information from the solar wind and geomagnetic activity and, importantly, their time histories. With a more capable representation of the organizing parameters and the target electron energy flux observations, PrecipNet achieves a $>$50\% reduction in errors from a current state-of-the-art model (OVATION Prime), better captures the dynamic changes of the auroral flux, and provides evidence that it can capably reconstruct mesoscale phenomena. We create and apply a new framework for space weather model evaluation that culminates previous guidance from across the solar-terrestrial research community. The research approach and results are representative of the `new frontier' of space weather research at the intersection of traditional and data science-driven discovery and provides a foundation for future efforts. 

\end{abstract}

%
%

%


%
%
%
%


\section{Plain Language Summary} \label{Plain Language Summary section}

Space weather is the impact of solar energy on society and a key to understanding it is the way that regions of space between the Sun and the Earth's surface are connected. One of the most important and most challenging to model are the way that energy is carried into the upper atmosphere (100-1000 km altitude). Particles moving along magnetic field lines `precipitate' into this region, carrying energy and momentum which drive space weather. We have produced a new model, using machine learning, that better captures the dynamics of this precipitation from a large volume of data. Machine learning models, carefully evaluated, are capable of better representing nonlinear relationships than simpler approaches. We reveal our approach to using machine learning for space weather and provide a new framework to understand these models.

\section{Introduction} \label{introduction section}
Electron precipitation is a key component linking the ionosphere and the magnetosphere. Electrons in the magnetosphere-ionosphere (MI) system carry current, transport energy, and precipitate (i.e., follow magnetic field lines from the magnetosphere to the ionosphere) to collide with the neutral atmosphere thereby driving changes in the electrical conductivity tensor. This tensor is central to the three dimensional electrical current circuit that flows over vast distances between the magnetosphere and the ionosphere. Indeed particle precipitation is a key input to all global circulation models (GCMs) such as the  Global Ionosphere Thermosphere Model (GITM) \cite{Ridley_2006}, the Thermosphere Ionosphere Electrodynamics General Circulation Model (TIE-GCM)  \cite{Roble_1988}, and the Whole Atmosphere Model-Ionosphere Plasmasphere Electrodynamics (WAM-IPE) model \cite{FullerRowell_1996, FullerRowell_2008} and is a starting point to the physical understanding of the ionosphere-thermosphere system. In a broader view, electron precipitation plays a role for space weather. Abrupt and intense precipitation events are associated with likewise abrupt and large amplitude changes in the ionospheric electron density and currents that drive ground induced currents. Such variability is linked to a number of potentially hazardous impacts, including errors in directional drilling, power grid transformer disruption, loss of Global Navigation Satellite Systems (GNSS) communication and corresponding timing and position accuracy \cite{Cannon_2013}. These are effects that need to be mitigated for the functioning of our increasingly technologically-dependent society. 

Models of electron precipitation and the associated ionospheric electrical conductivity tensor elements have been proposed over the last few decades (for a review of such models, see \citeA{Newell_2014}, \citeA{Machol_2012}, or Chapter 2 of \citeA{McGranaghan_2016}). In Section \ref{particle precipitation modeling section} we provide a brief overview of those models and the details relevant to the model we develop as part of the present study. These models produce 2D spatial distributions of particle energy and number flux as a function of some assumed driving parameters. They are generally limited to the northern hemisphere and rely on limited and instantaneous organizing/input parameters. For example, the highly successful and widely used \citeA{Hardy_1985} model is driven by the planetary Kp magnetic index whose time resolution is three hours and prevents the model from capturing changes that occur on a shorter time-scales, such as substorms. More recently, the series of Oval Variation, Assessment, Tracking, Intensity, and Online Nowcasting (OVATION) models \cite{Newell_2010a, Newell_2010b, Newell_2010c} are driven by a different set of input data such as solar wind parameters and geomagnetic indices. The OVATION models produced a different paradigm of particle precipitation specification, one that has brought corresponding change in capabilities. OVATION uses solar wind parameters, making the implicit assumption that these parameters are the cause and the observed electron precipitation is the effect. Moreover, analysis of data from geospace (e.g., geomagnetic indices) and from LEO satellites shows that these parameters provide information of the simultaneous and global electron precipitation distribution (e.g., `state descriptors' \cite{Gjerloev_2018}). The success and adoption of the OVATION models (one of the most downloaded products from the National Oceanic and Atmospheric Administration (NOAA) Space Weather Prediction Center (SWPC) website \url{https://www.swpc.noaa.gov/products/aurora-30-minute-forecast}) give credence to these assumptions and allude to a central point of this paper that rich information exists in the spectrum of Heliophysics observational data, and that careful representation of this information can lead to profound progress. 

However, existing models are ill-equipped to provide the critical information to space weather users and to inform physics-based models and understanding. Much of the limitation is the result of shortcomings in information representation (in the form of using a limited set of state descriptors or creating representations that do not capture the complexity, e.g. nonlinearity, of the system). In the particle precipitation application, such shortcomings have resulted in models that are limited in their ability to reproduce observed features in highly dynamic (in both space and time) conditions. However, such features are produced during the frequent phenomenon of magnetosphere-ionosphere (MI) system reconfiguration and thus the quality of any model is limited if the modeled variations in time and space are smooth.

Our development uses methods from the burgeoning field of machine learning (ML), made possible through new capabilities in the management, processing, analysis, and communication of data (i.e., data science) \cite{McGranaghan_2017}, to increase the utility of the available data through identification of more representative relationships between potential drivers and particle precipitation. More explicitly, we examine the extent to which neural network ML methods can more capably capture complex nonlinear relationships between input (e.g., solar wind conditions and their time history) and output (total electron energy flux) than previous parameterizations which rely on lower-dimensional relationships. 

ML is often a misunderstood term, so it is important to explicitly identify our usage. We define ML in a broad sense, encompassing any approach that allows a computer system to learn from experience introduced in the form of data samples \cite{Mitchell_1997}. This definition encapsulates a broad range of approaches, including linear regression, clustering, information theory, statistical modeling, and neural networks, to name a few. Timing is ripe to advance particle precipitation models through data science and ML for three reasons: (1) the availability of a large volume of high-quality data, (2) an understanding of well-defined improvements to current models that are required to serve scientific and operational needs and a recognition of precisely how ML could be used to deliver such improvements, and (3) accessible adequate computational resources. Full reviews of such models have been given elsewhere, e.g., \citeA{Machol_2012, Newell_2015, McGranaghan_2016}.

We present a new nowcast model of total electron energy flux based on LEO measurements of electron precipitation, namely the Defense Meteorological Satellite Program (DMSP) satellites \url{https://cdaweb.gsfc.nasa.gov/pub/data/dmsp/}; \cite{Redmon_2017}, driven by a combination of solar wind parameters and state descriptors and their time histories. The advantage of this combination is that state descriptors and solar wind parameters allow for detailed nowcasting solutions while the model also allows for forecasting with an increasing dependence on solar wind parameters as the forecasting extends into the future. We present the nowcasting results in this manuscript. Recognizing the need to more accurately represent the information in these data, we explore more capable (e.g., nonlinear) relationships between the input solar wind and state descriptors and output particle precipitation via ML. We do \textit{not} address the myriad important operational considerations, such as various data latencies that must be considered to stream real-time data, in this first description of our new model.

Our approach is to align ML with an understanding of the physical system and robustly `interrogate' the resultant ML model. We carry out a comprehensive interrogation of the model at several levels. We hope that this approach will be an example for future ML investigations to better incorporate physics, provide greater insight to the ML model, and to produce more fruitful discovery within a `New Frontier' at the intersection of traditional approaches and state-of-the-art data-driven sciences and technologies \cite{McGranaghan_2017}.

In terms of improving existing models, we address the following critical needs: 
\begin{enumerate}
    \item Utilize diverse observations: We develop a baseline model using DMSP data, but robustly interrogate using a wide variety of information, including auroral imagery data. Further, our approach is naturally extensible to any observation that provides particle energy fluxes; 
    \item Establish more informative model inputs: In the evolution of precipitation models it is clear that the input parameters dictated improvements and differences in capability. Moving from less (e.g., Kp) to more capable (e.g., the \citeA{Newell_2007} Coupling Function (NCF)) drivers allowed commensurate improvement in the specification of precipitation. We embrace this development and use ML to capture more capable (i.e., higher dimensional and with higher nonlinearity) drivers of the precipitation. In Section \ref{feature importance subsection} we quantify the importance of the drivers and state descriptors for precipitation; and
    \item Capture finer-scale variations in space and time: Existing models are limited in their ability to reproduce observed features that are associated with \change{large}{steep} spatial gradients and that occur rapidly.  Such features, however, are produced as part of the frequent reconfiguration of the MI system (on average three times per day). Scientific and operational needs dictate specification at mesoscales (defined here to be 100s of km and 10s of minutes). Therefore, mesoscale and substorm-level spatial temporal specification is our context in this work. 
\end{enumerate}
We present a new nowcast model of total electron energy flux trained on DMSP measurements of electron precipitation. The model is driven by a combination of solar wind parameters and state descriptors. The remainder of the paper is laid out as follows: Section \ref{data landscape section} details the landscape of the data supporting particle precipitation specification and our use of the opportunities afforded by it; Section \ref{particle precipitation modeling section} reviews the existing particle precipitation models, focusing on the OVATION models, and links the science needs to the use of an ML approach; Section \ref{model exploration section} reveals our ML exploration process and the model discovered; and Sections \ref{interrogation section} and \ref{conclusion section} examine the model performance, discuss the findings, and conclude the paper.

\section{The Heliophysics and Space Weather data landscape in the context of particle precipitation} \label{data landscape section}

\begin{figure}[h] 
\caption{Visual representation of the machine learning (i.e., analysis ready) data preparation steps for an illustrative time period containing substorm behavior on February 6, 2002 when DMSP F13 passed over mesoscale undulations around 0857 UT. (a) Polar plots of the DMSP F16 satellite northern hemisphere pass with one-second (left) and one-minute sub-sampled (right) total electron energy flux data. The green ovals circle indicate a period explored further in subplot (c). Data are shown in AACGM MLAT-MLT coordinates with noon MLT to the top of each polar plot and a low-latitude limit of 50$^{\circ}$. AACGM = altitude adjusted geomagnetic coordinates; MLAT = magnetic latitude; MLT = magnetic local time. (b) Time series data for selected organizing parameters (the interplanetary magnetic field (IMF, top), SymH index (second row), and AE index (third row) corresponding to a given total electron energy flux (bottom row) for February 6, 2002. The vertical pink bar indicates a mesoscale undulation on the dusk side during the substorm expansion/recovery phase. Green shaded area corresponds to the green ovals in subplot (a). (c) Zoomed time series for green shaded period in (b), revealing that the sub-sampled DMSP data maintain information about mesoscale activity like substorms (the pink arrow points out the spike in observations corresponding to the undulation). (d) Visual demonstrating the calculation of time histories of each organizing parameter, shown here for the AE index where grey shaded regions and vertical bars represent distinct time history points at 6, 3, and 1 h prior to $t=0$ as well as 45, 30, 15, 10, and 5 min prior to $t=0$. (e) Schematic of construction of the analysis ready data samples consisting of the organizing parameters and the corresponding DMSP total electron energy flux at a given time step as one row in the matrix.}
\label{data preparation figure}
\centering
\includegraphics[width=0.6\textwidth]{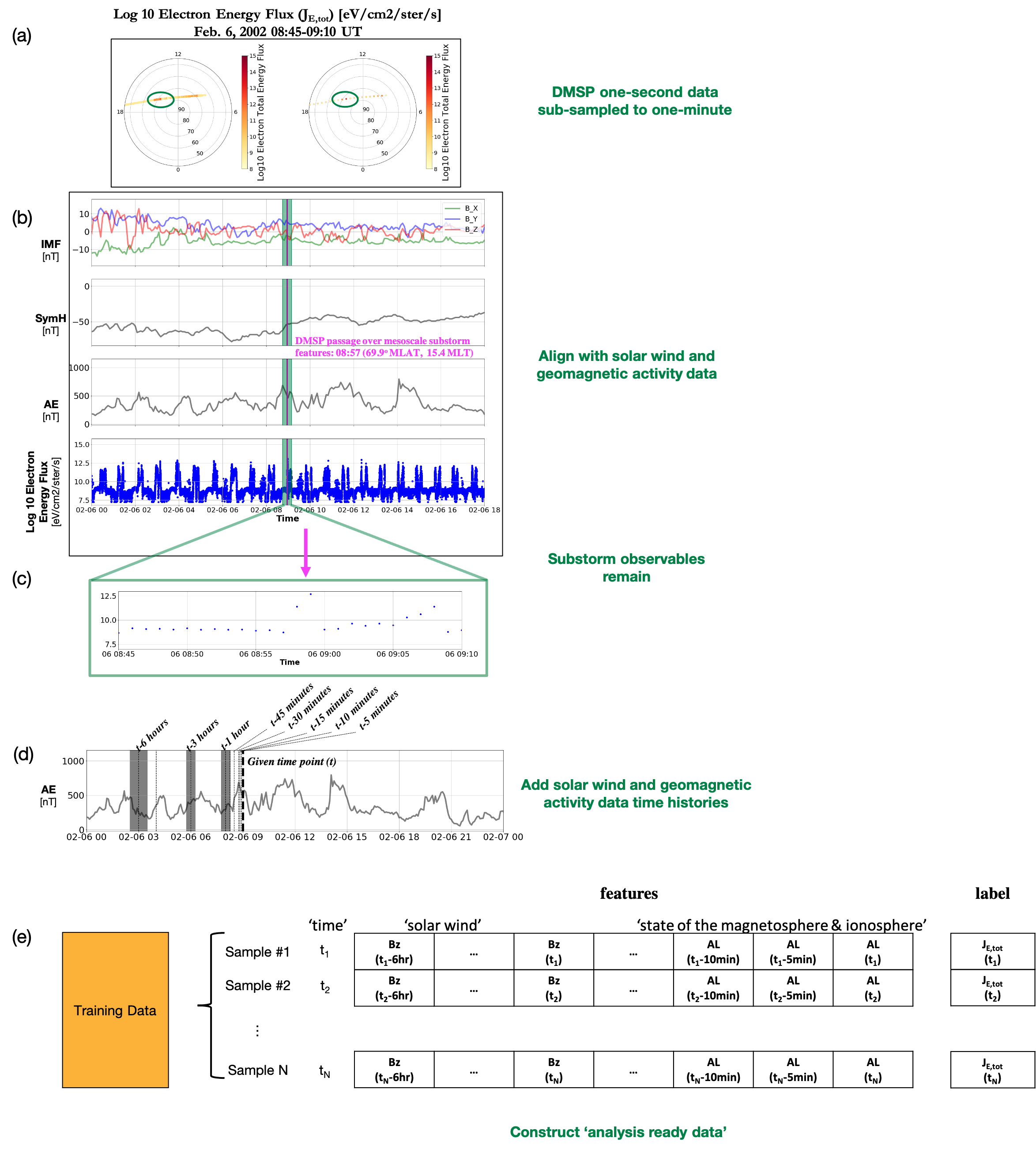}
\end{figure}

The expanse of the Heliophysics system makes it unrealistic to deterministically observe all components of the solar-terrestrial system simultaneously. Rather, our understanding relies on ingenuity in the combination of heterogeneous data sets; the problem is particularly important to particle precipitation, which must link solar wind drivers to disparate representations of the magnetospheric and ionospheric states. Here we provide detailed information about our preparation of the data (i.e., our data pipeline from the existing data archives to an `analysis ready' database \cite{Ramachandran_2018}) to: 1) give transparency to our methods and promote reproducibility and 2) to create a `challenge data set' published here: \citeA{McGranaghan_challenge_data_set}) around which this work can be expanded. Our data preparation is visually represented using an illustrative time period containing substorm behavior on February 6, 2002 when DMSP F13 passed over mesoscale undulations around 0857 UT presented in Figure \ref{data preparation figure}. \add{We emphasize that} Figure \ref{data preparation figure} \add{is not intended to be definitive, but simply to illustrate the complex relationship between solar wind and geomagnetic indices and particle precipitation and that our data preparation was carefully designed to allow our database to be representative of those relationships. }This pass was chosen because F13 observed substorm-related mesoscale phenomena as analyzed by \citeA{Lewis_2005} and illustrates that our data processing both retains substorm-scale observables and contains the Heliophysics system information needed to characterize such mesoscale behavior. The explicit data preparation steps are: 
\begin{enumerate}[label=(\alph*)]
    \item Obtain DMSP one-second data and \textit{sub-sample} (note it is important that we do not average) to a one-minute cadence from the NASA Coordinated Data Analysis Web (CDAWeb; \url{https://cdaweb.sci.gsfc.nasa.gov/index.html/}) system. The left polar plot of Figure \ref{data preparation figure}(a) shows the DMSP F13 spacecraft during the period 0845-0910 at one-second cadence, while the right shows the same trajectory, though with the one-minute sub-sampled data points. Colors represent the total electron energy flux values [$\frac{eV}{cm^2 \cdot ster \cdot s}$] on a log10 scale. This step includes a robust data quality check using flags in the DMSP data \cite{Redmon_2017}. The green rings encircle a portion of the trajectory in which substorm behavior is observed that is explored in Figure \ref{data preparation figure}c; 
    \item Align with solar wind and geomagnetic indices (i.e., ionosphere and magnetosphere state descriptors) data. For each time stamp of the DMSP data we attach the solar wind and geomagnetic activity indices data. Figure \ref{data preparation figure}(b) shows the data for an illustrative subset (from top to bottom: solar wind interplanetary magnetic field (IMF, GSM coordinates), Symmetric-H (SymH) index, Auroral Electrojet (AE) index, and the DMSP total electron energy flux observation). The geomagnetic activity indices are used as ionosphere and magnetosphere state descriptors \cite{Borovsky_2018}. Again, we use flags to identify missing or questionable data as detailed by the CDAWeb system (\url{https://cdaweb.gsfc.nasa.gov/pub/000_readme.txt}). Note that in the OMNI data within CDAWeb (\url{https://omniweb.gsfc.nasa.gov/ow_min.html}: ``High resolution (1-min, 5-min) OMNI: Solar wind magnetic field and plasma data at Earth's Bow Shock Nose (BSN)'') the solar wind has been propagated to the magnetopause location and that we use these propagated data. The portion of the trajectory in Figure \ref{data preparation figure}(a) that is circled in green is shaded green on this panel. The substorm identified by \citeA{Lewis_2005} is explicitly identified by the vertical magenta line and in the corresponding magenta text. 
    \item Demonstrate that information on substorm activity is observed in the DMSP data. We zoom in on the shaded green portion of the time period to reveal that the sub-sampled DMSP total electron energy fluxes experience enhancements of two orders of magnitude during its passage of the substorm activity that permeates this short time span. The enhancement is representative of those observed in the one-second data and analyzed by \citeA{Lewis_2005}. Moreover, looking at Figure \ref{data preparation figure}b, it is clear that the solar wind and selected geomagnetic activity indices data reveal correspondence.  
    \item Add solar wind and geomagnetic activity data time histories. In step (b) we aligned the instantaneous features (solar wind and geomagnetic activity indices) to the DMSP observation data samples, but recognize that previous information from the solar wind and ionosphere and magnetosphere are also important to current particle precipitation and substorm behavior \cite[and references therein]{Borovsky_2020b}. In order to capture such information we attach to the data samples the time histories of each of the solar wind and geomagnetic activity indices data. This process is detailed in Figure \ref{data preparation figure}(d) for the AE index during the time period chosen. Previous approaches have for simplicity assumed that the instantaneous values of these data points are sufficient, however we recognize that the purpose of these time history data are different for near and far periods. For instance, over shorter time scales (minutes to $\sim$an hour) solar wind time history provides a snapshot of current conditions, while over longer time scales (from hours to years) the solar wind time history provides climatology or trend information. Therefore, we treat the previous time steps in two different ways: for all data points within one hour of the DMSP observation we include the instantaneous value and beyond that we gradually increase number of data points centered on the $t-x$ point over which an average is calculated, where $x$ is any number of minutes before the time of the DMSP observation--for the $t-$1 h and $t-$3 h points we use a 30-minute average and for the $t-$6 h point we use a one-hour average. Together with the instantaneous aligned data shown in Figure \ref{data preparation figure}(c), the complete data set contains comprehensive information to specify and predict substorm-scale phenomena.
    \item Construct `analysis-ready data' \cite{Ramachandran_2018}. Figure \ref{data preparation figure}(e) shows data samples construction, pulling from each of the panels above, utilizing `tidy data' guidelines considered a best practice in the data science community \cite{Wickham_2014}. The outcome is data samples that each contain the driver (solar wind), state descriptors of the ionosphere and magnetosphere, and resultant DMSP observation. Together, the set of input data are referred to as `features.' In total more than 150 features are included--the full list is provided at \url{https://github.com/rmcgranaghan/precipNet}. The high dimensionality of these data both reveal the amount of information required to capture substorm-scale behavior in a particle precipitation model and motivate machine learning approaches as models capable of the requisite `expressive power.' It is trivial to expand the data used for this model or other ML models by preparing any data set that provides energy flux data in this analysis-ready format. 
\end{enumerate}

Sub-sampling the one-second DMSP observations both reduces the data volume, which was important to computational feasibility for the expansive ML optimization process we applied (see Section \ref{model choice subsection}), and improves the assumption that we make that DMSP observations are independent. 

Two final points about data preparation. First, we remove observations that represent unrealistically large fluxes, deemed to be those exceeding the 99.995th percentile (equal to $7.37\times 10^{13}$ eV/cm$^2$/sr/s or $\sim$118 erg/cm$^2$/sr/s), which was empirically determined to separate values that were nonphysical. Second, we combine the northern and southern hemispheres. The sun-synchronous orbits of the DMSP spacecraft restricts their MLT coverage, necessitating the combination of northern and southern hemisphere observations to complete the local time coverage. This point has been well-covered in previous works (e.g., \citeA{Newell_2015}) and we do not belabor it here. However, we do attempt to combine the data according to recent understanding of hemispheric asymmetries \cite{Laundal_2017}. Namely, the signs of the IMF B$_{\textrm{y}}$ for the southern hemisphere data are reversed to align with the northern hemisphere data (\cite{Laundal_2017} and J. P. Reistad private communication). We have not altered the southern hemisphere data to account for local season, however, which may affect the model's ability to capture seasonal variations. 

The full database consists of 17 distinct years (1987--1988, 2000--2014) and 51 satellite years (DMSP generally has three operational satellites at any given time, which includes F6-F8 and F12-F18), constituting more than 1.9 million observations. In terms of the calibration of these instruments, the DMSP spacecraft each undergo an in‐flight calibration (IFC) procedure \cite{Emery_2006}. Note that the DMSP spacecraft began carrying a different version of the particle spectrograph instrument (SSJ/5) from the F16 satellite onward such that observations used in our database consist of both SSJ/4 (pre-F16) and SSJ/5 (post-F16). As a means of intersatellite calibration the average IFC factor across all SSJ/4 and SSJ/5 instruments is used as a reference with which to normalize the individual IFC factors (R. Redmon and E. Holeman, personal communication, 2015) \cite{McGranaghan_2015}. More information on the processing of DMSP data is provided by \citeA{Redmon_2017}. We use the publicly available data and do not make further adjustments between satellites. As a final note on the SSJ instruments, the version of the DMSP data from CDAWeb that we have processed includes upgrades to the calibration factors to account for instrument degradation over time \cite{Redmon_2017}.

\begin{figure}[h] 
\caption{(Caption on next page.)}
\label{data coverage figure}
\centering
\includegraphics[width=0.6\textwidth]{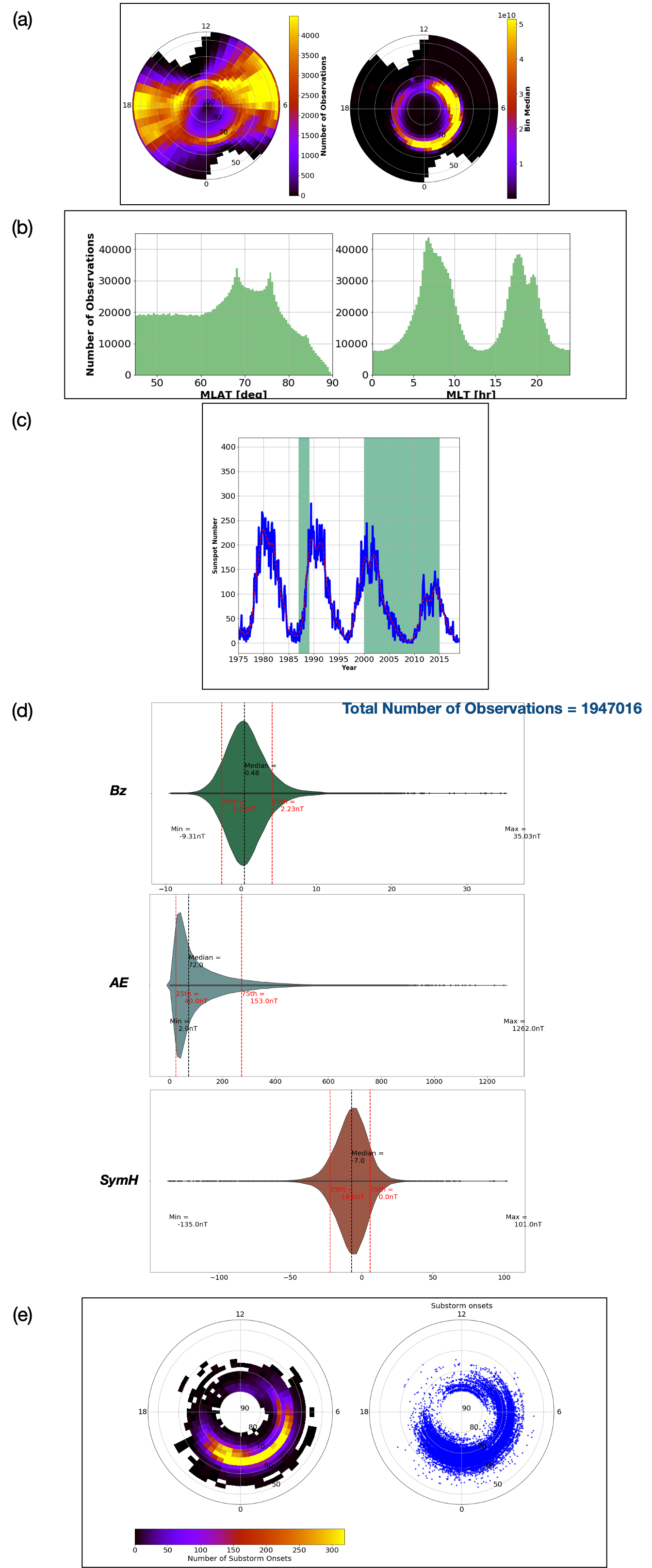}
\end{figure}

We attempt to expose any biases in the data to understand the capabilities and limitations of the resultant model. Figure \ref{data coverage figure} shows the coverage provided by the data in multiple dimensions: (a) geomagnetic latitude; (b) magnetic local time; (c) solar activity; (d) geomagnetic activity; and (e) substorm occurrence (our context in this work). 

\addtocounter{figure}{-1}
\begin{figure} [t!]
\caption{Complete data coverage details: (a) MLAT-MLT observational density (left panel) and total electron energy flux binned median values (right panel) in AACGM MLAT-MLT coordinates with noon MLT to the top and a low-latitude limit of 50$^{\circ}$ at a resolution of 0.5 hours MLT and 2$^{\circ}$ MLAT. AACGM = altitude adjusted geomagnetic coordinates; MLAT = magnetic latitude; MLT = magnetic local time. (b) Histograms of observational density broken out by MLAT (left panel) and MLT (right panel). (c) Universal scale contextual perspective: monthly mean sunspot numbers. The blue trace shows the monthly mean numbers themselves, while the dashed red trace are the 13-month smoothed values. The portions of the plot shaded green are those for which observational data were used in this work. (d) Global scale contextual perspective: geomagnetic activity relevant to particle precipitation represented by B$_{\textrm{z}}$, AE index, and $Sym$-$H$ index. Each parameter is represented using a violin plot that reveal the lower and upper quartiles (vertical red dashed lines and accompanying text), the median (vertical black dashed line and accompanying text), labels for the minimum and maximum to provide a sense of extrema, and each individual data point as small black scatter points to show the full range. The plot also shows the full distribution of the data mirrored across the center line. Wider sections of the violin plot represent a higher probability of observations taking a given value, the thinner sections correspond to a lower probability. (e) Local scale contextual perspective: substorm onsets in the SuperMAG database for the time periods covered by our database. Both the scatter (right panel) and density (left panel) polar plots are shown. SuperMAG = Super Magnetometer Initiative.}
\end{figure}

Polar plots in Figure \ref{data coverage figure}(a) reveal the MLAT-MLT observational density and total electron energy flux binned median values in altitude adjusted geomagnetic coordinates (AACGM) magnetic latitude (MLAT)-magnetic local time (MLT) coordinates with noon MLT to the top and a low-latitude limit of 50$^{\circ}$. A resolution of 0.5 hours MLT and 2$^{\circ}$ MLAT, sufficient to reveal gross spatial region biases in the database, e.g., whether the dawnside contained more observations than the duskside, was chosen for these figures. To augment the observational density polar plot (left panel) Figure \ref{data coverage figure}(b) shows histograms broken out by MLAT (left) and MLT (right). These figures reveal that there are natural, but limited, spatial biases in the DMSP observational database used for this study. The MLAT histogram shows minor peaks around 67 and 76$^{\circ}$, which are due to the various orbits from the F6-F18 missions with respect to the geomagnetic pole across the 1987-1988 and 2000-2014. There is a slight decrease in the number of data points poleward of 80$^{\circ}$ MLAT. The two more prominent peaks in the dawn and dusk MLT sectors (right panel of Figure \ref{data coverage figure}(b)) reveal the inevitable impact of the DMSP sun-synchronous orbits and predominance of dawn-dusk orbital trajectories of the spacecraft. The left polar plot of Figure \ref{data coverage figure}(a) spatially illustrates these trends. The median values of the total electron energy flux in the right panel of Figure \ref{data coverage figure}(a) shows that unsurprisingly a strong signature of the auroral oval emerges in the precipitation data. Median values are shown to reduce the effect of extreme values, though disturbed oval activity is apparent in heightened values extending from $\sim 64^{\circ}$ to greater than $\sim 72^{\circ}$ MLAT in the midnight and post-midnight MLT sectors and in the dusk and post-dawn MLT sectors, areas of predominantly discrete precipitation that is associated with geomagnetic activity \cite{Newell_2010a}. 

Figures \ref{data coverage figure}(c)--(e) provide contextual coverage information for the DMSP database, beginning at the largest scale (the sun) and driving to finest scales in this work (substorms). Figure \ref{data coverage figure}(c) shows the monthly mean sunspot numbers obtained and freely available from the Royal Observatory of Belgium, Brussels and their online Sunspot Index and Long-term Solar Observations (SILSO) service (\url{http://www.sidc.be/silso/home}). The blue trace shows the monthly mean sunspot numbers themselves, while the dashed red trace are the 13-month smoothed values. The portions of the plot shaded green are those for which observational data were used in this work. Those data cover 17 years, including two full cycles of ascending cycle (1987-1988 and 2009-2014), the full solar cycle from the maximum of solar cycle 23 (in March 2000) to the maximum of solar cycle 24 (in April 2014), and portions during solar minimum. Thus, the solar cycle is well sampled by our database. 

Next, we examine the coverage at the global scale and as a function of geomagnetic activity. In Figure \ref{data coverage figure}(d) we emulate the informative presentation of geomagnetic coverage shown in Figures 1 and 2 of \citeA{Welling_2017} because it is a direct representation of the range of information over which a model will be useful and therefore serves the same purpose in both manuscripts. The B$_{\textrm{z}}$, AE index, and $Sym$-$H$ index are chosen to represent the activity of relevance to particle precipitation. Each parameter is represented using a violin plot that reveals the lower and upper quartiles (vertical dashed red lines and accompanying text), the median (vertical dashed black line and accompanying text), text labels for the extrema, and each individual data point as small black points to show the full range. The solid colored regions are the full distribution of the data mirrored over the center line. The median values (B$_{\textrm{z}}$=0.48 nT, AE=72.0 nT, $Sym$-$H$=-7.0 nT) indicate that quiet conditions are the predominant behavior. However, the width of the violin plots show the breadth of conditions covered by the database, covering B$_{\textrm{z}}$ conditions between -10 and 35 nT, AE between 2 and $>$1250 nT, and $Sym$-$H$ reaching below -130 nT. The ranges provide confidence that our model contains information from a robust variety of conditions. Although we show values for the full database, the values are essentially unchanged when isolating the training data (those used to develop the model and over which the model can be considered `valid'). We acknowledge that the DMSP database is B$_{\textrm{y}}$-dominant \cite{Newell_2004}, meaning that the typical magnitude of B$_{\textrm{y}}$, the component of the IMF that is perpendicular to the Earth-Sun (X) and projection of Earth's dipole (Z) on the plane perpendicular to X directions, is significantly larger than the typical magnitude of B$_{\textrm{z}}$ due to the predominance of the IMF to lie in the ecliptic. 

Figure \ref{data coverage figure}(e) examines the local or substorm-scale representativeness of our database. We obtained the list of all substorm onsets in the Super Magnetometer Initiative (SuperMAG) database \cite{Newell_2011, Newell_2011b} for the time periods covered by the DMSP observations in our database and show both the scatter (right panel) and density (left panel) polar plots. The onsets reveal that peaks in substorm onset occur in the 22-24 MLT sector between $\sim66-72^{\circ}$ MLAT, in good agreement with previous analyses \cite[and references therein]{Gjerloev_2007}. However, substorm onsets are observed over a much more broad range of MLTs and MLATs and comparison with Figure \ref{data coverage figure}(a) shows that these locations are well-sampled by the DMSP database. 

Software to reproduce analyses in this manuscript and to extend the research are provided at \url{https://github.com/rmcgranaghan/precipNet}.

\section{Particle precipitation modeling} \label{particle precipitation modeling section}

We now briefly review a few of the most widely used auroral particle precipitation models, focusing on the needs that drive the development of a new model. Full reviews of such models have been given elsewhere (e.g., \citeA{Machol_2012, Newell_2015, McGranaghan_2016}. These models are:
\begin{enumerate}
    \item \citeA{Hardy_1985, Hardy_1987}: Observations of auroral electron energy and number flux from the DMSP F2 and F4 spacecraft were compiled and used to produce statistical patterns on a 2$^{\circ}$ latitude-0.5 hour local time grid parameterized by the Kp index;
    \item \citeA{FullerRowell_1987}: Observations of energy flux and characteristic energy from the NOAA POES spacecraft were compiled and used to produce statistical patterns on a 1$^{\circ}$ latitude-2$^{\circ}$ ($\sim0.13$ hour) local time grid parameterized by the hemispheric power index; 
    \item \citeA{Newell_2010a}: The Oval Variation, Assessment, Tracking, Intensity, and Online Nowcasting (OVATION): Observations of auroral electron energy and number flux from all DMSP spacecraft between 1988 and 1998 were compiled and used to produce statistical patterns parameterized by latitude (0.5$^{\circ}$), local time (0.25 hour bins), and the NCF \cite{Newell_2007}. OVATION has since become OVATION Prime where the `Prime' refers to an upgrade of the original model to provide information about the different types of contributors from diffuse, monoenergetic, and wave/broadband aurora. 
\end{enumerate}

\citeA{Newell_2015} provide a thorough breakdown of \textit{types} of auroral particle precipitation models, principally categorized by their organizing parameters (solar wind, geomagnetic indices, or direct in-situ particle precipitation observations). That work makes useful suggestions of needed future developments, many of which trace to a core need to \textit{better utilize more data}. We briefly enumerate the salient suggestions from that work:
\begin{enumerate}
    \item The organizing parameters of auroral particle precipitation models most strongly determine their predictive ability and utility; 
    \item Simple binning of data does not produce capable results; and 
    \item Direct in-situ observations of the precipitation are invaluable to precipitation model development.
\end{enumerate}

Our machine learning approach (see Section \ref{why ml subsection}) responds to these needs, using information from the solar wind, geomagnetic indices, and direct observations to construct a new model.


\subsection{Oval Variation, Assessment, Tracking, Intensity, and Online Nowcasting (OVATION) model}
The OVATION particle precipitation model was introduced by \citeA{Newell_2002} as a means to locate the auroral oval and specify the intensity of the auroral particle precipitation. The original version was developed for nowcast operational use and shares little resemblance to the later versions most often discussed in the literature. Later versions, called `OVATION Prime' models, serve the same purpose but are solar wind-driven \cite{Newell_2009}, include seasonal variations \citeA{Newell_2010a}, and distinguish the various types of particle precipitation. The first OVATION Prime is described in \citeA{Newell_2010a} and is the version of the model that we use for all work in this manuscript. The source code to run OVATION Prime is available from Janet Machol, Rob Redmon, and Nathan Case of the National Oceanic and Atmospheric Administration (NOAA) National Center for Environmental Information (NCEI) on Sourceforge (\url{https://sourceforge.net/projects/ovation-prime/}) and a Python implementation from Liam Kilcommons on GitHub (\url{https://github.com/lkilcommxons/OvationPyme}). We used the Python implementation and for purposes of reproducibility have tagged the specific Github commit of OvationPyme used for this manuscript, which can be obtained from \url{https://github.com/lkilcommons/OvationPyme/releases/tag/v0.1.1}. This is also detailed in the PrecipNet Github repository. It is important to note two other variations on the model that exist: 1) an extension of the base model to better specify precipitation for Kp$>$6 using auroral imagery data from the Global Ultraviolet Imager (GUVI) on the TIMED (Thermosphere, Ionosphere, Mesosphere Energetics and Dynamics) satellite which was labeled OVATION Prime 2013; and 2) \citeA{Mitchell_2013} a different approach based on ground magnetometer data and fitting to a generalized auroral electrojet (AE) index, while using the DMSP particle precipitation observations to separate precipitation type and was called OVATION-SuperMAG (SM). These versions are not readily available and the operational model in use at NOAA SWPC is the 2010 version which motivated our use of that version. Extensive discussion, including clear distinction between versions, is provided in \citeA{Newell_2014} (for OVATION Prime 2010 vs. 2013) and \citeA{Newell_2015} (for precipitation model types, in general). 

We consider OVATION Prime to be the `state-of-the-art', and therefore the most appropriate focus of our comparisons, due to its high performance across a range of studies and validation efforts, using various criteria, and wide application \cite{Newell_2010b, Machol_2012}. 

\subsection{Why a machine learning model?} \label{why ml subsection}


\citeA{Newell_2014} observe that to date most precipitation models have relied on a small number of discrete `bins' for the parameters that are used to drive the models (e.g., a Kp-based model such as \citeA{Hardy_1985} provides precipitation `maps' for each level of the 1-9 index), leading to a few discrete possible patterns with large jumps in between. The notable exception is OVATION which calculated linear regression fits of the auroral electron energy and number fluxes parameterized by the NCF \cite{Newell_2007}, representing an advance over the more discrete approaches and reaping concomitant advances \cite[and references therein]{Machol_2012, Newell_2014}. Considering the regression fit approach as an expansion of model `expressive power,' the ability of the model to represent observations, OVATION demonstrated that progress could be achieved through expansion of model capability through a more robust treatment of the input parameters (e.g., the NCF) and constructing a more continuous representation (e.g., regression fits instead of Kp parameterization). 

Therefore, future advances are likely to come from similar expansions and ML represents a set of methodologies for precisely this context. Indeed ML provides an algorithmic means to incorporate higher dimensional input parameter spaces (discussed in Section \ref{feature importance subsection} below) and create models with greater expressive power. These possibilities come with their own challenges, such as overfitting (see, for instance, Chapter 11 of \citeA{Hastie_2001}), which must be carefully addressed. Below we comprehensively introduce these design challenges and our approach through neural networks to produce models that reap the benefit of ML and attempt to overcome their challenges in a physics-informed and data-driven manner. 

A final note of importance to the value of the ML approach is that OVATION relies on bins in latitude and local time, separately fitting a linear model in each bin. Therefore, the model produces discontinuous outputs across bins in a 2D map, while the ML approach, by design, produces 2D output continuous across grid points. 

\section{Methodology to address existing shortcomings in particle precipitation modeling} \label{model exploration section}
In this section, we identify the design space for ML studies and the specific experiments we used to make data-driven and physics-informed decisions across it. Our results are presented for the specification (i.e., \textit{nowcast}) of total electron energy flux.

Given open questions about which variables are most important to the specification and prediction of particle precipitation \cite{Newell_2007, Newell_2011, Newell_2015}, we give specific attention to this topic and then use these findings to provide details of the ML model developed. 

\subsection{What input `features' are most important?} \label{feature importance subsection}

In ML parlance, `features' refers to the variables passed as input to the model. For instance, the feature for the OVATION models is the NCF (and, implicitly, the solar wind IMF and velocity). Choice of features represents one of the most apparent ways by which physical understanding can inform ML model development. We recognized the need to expand the feature set and therefore have compiled a set of 153 input features (the full set of which can be found in the repository for this work: \url{https://github.com/rmcgranaghan/precipNet}), including the time history of features as individual features. These features were selected based on the likely existence of a physical relationship with high-latitude particle precipitation (see, for instance, the important discriminating factors identified by \citeA{Ridley_2004, Newell_2007, Lu_2016, Borovsky_2018} for high-latitude particle precipitation). Additionally, past studies have shown that solar wind information alone is insufficient to characterize auroral behavior (e.g., changes in critical auroral parameters such as hemispheric power (HP) on substorm temporal scales (10s minutes) \cite{Newell_2001, Newell_2010c}) and that quantification of the magnetospheric state and monitoring the global activity level are required for models that hope to capture substorm-scale phenomena \cite{Richmond_1998, Liou_1998, Newell_2010b, Nishimura_2016, Deng_2019}. These considerations drove the inclusive set of features that we explore in this work, compiling information from the solar wind (e.g, IMF B$_{\textrm{z}}$), throughout the magnetosphere (e.g., the Borovsky coupling function), and the global activity level (e.g., the symmetric-H ($Sym$-$H$) index). Note that \remove{determination of the first expansive set of features (the full 153 feature set) was achieved through extensive discussion and deliberation by a team chosen for their breadth of knowledge about the particle precipitation problem} \remove{Therefore, }prior to the algorithmic feature importance exploration, first physical understanding was used to construct the possible features of use (this occurred in an International Space Sciences Institute (ISSI) working group \cite{Mcgranaghan_2019}). 

\begin{figure}[h] 
\caption{Feature importance results for: (a) Principal Component Analysis (PCA); and (b) Random Forest (RF). The y-axis is the magnitude of the importance and the x-axis is the time point of the parameter (i.e., `6hr' refers to the data points six hours prior to the DMSP observation for each feature). Each parameter is given a unique color that is consistent across both subplots and the corresponding textual label is placed at the time point of maximum importance. (c) A table showing the top 15 most important features from our analysis.}
\label{feature selection results figure}
\centering
\includegraphics[width=0.6\textwidth]{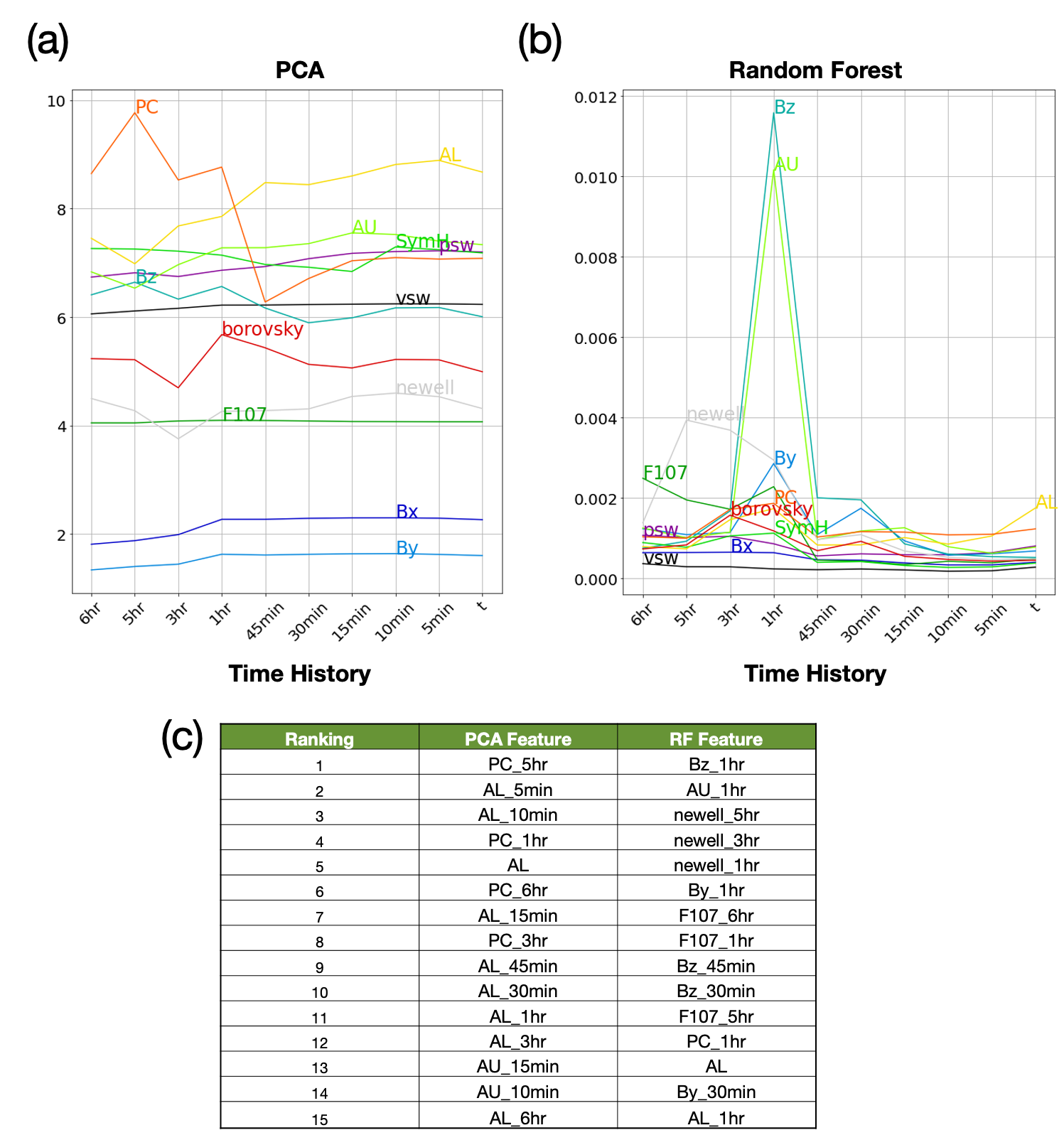}
\end{figure}

However, higher dimensionality of input features (i.e., selecting more input features) may result in lower performance. Indeed, intercorrelations between the various solar wind features and intercorrelations between various geomagnetic activity indices mean that numerous features may carry information about other features and together be redundant \citeA[and references therein]{Borovsky_2020, Bentley_2018}. Performance improvement can be gained by removing less informative features. \remove{This process is typically called featurization and is an active area of research in the machine learning community. Featurization not only results in a smaller set of input features, and, therefore, a simpler model, but can also improve physical understanding by quantifying the relationships between the inputs and the predicted variable.} We used two feature importance algorithms to examine which of the inputs emerge as most informative. Figure \ref{feature selection results figure} shows the results from: (a) principal component analysis (PCA) \cite{jolliffe_cadima_2016}; and (b) random forest (RF) \cite{Breiman_2001}. PCA linearly transforms the input variables by rotating them to align with their principal components, i.e. new variables that are constructed as linear combinations or mixtures of the initial variables that represent the orthogonal directions of maximum variance. These directions are defined by the eigenvectors of the input features covariance matrix. In PCA, the importance is obtained by the magnitude of the values corresponding to a given feature in the eigenvector, and the final importance is the sum of the absolute value of these across the eigenvectors that together capture $>$90\% variability in the DMSP observations. RF is a collection of decision trees, which operate by iteratively splitting data samples into two branches based on one input feature to attempt to create two sets from one where each new set has increased similarity and is distinct from the other split. Thus, the importance of a given input feature can be quantified by the amount by which variances in the split sets are reduced. With the RF approach, we average the importance for each input feature for many decision trees (100 in this work). Thus, both algorithms provide some measure of the relationship between an input feature and the target. Neither are perfect measures, but together they provide useful information when combined with an initial set of physically-informed features. Note that we performed all feature importance analyses on non-normalized input features. We used ten-fold cross-validation (CV) \cite{Hastie_2001} to examine ten separate cases of input data and plot the median of these ten feature importance calculations. In Figure \ref{feature selection results figure} we have removed features identified as redundant (V$_\textrm{sw,x}$ and AE, whose information are fairly robustly encoded in V$_\textrm{sw}$ and AU/AL). Using these results we can also determine the important time history steps to keep (note that the y-axis in the figure refers to the time history point for a given input feature such that the plotted point for `B$_\textrm{Z}$' at `3hr' indicates the importance of B$_\textrm{Z}$ three hours before the DMSP observation). It is perhaps meaningful to note that there is little overlap between the top 15 features in the PCA and RF importances (with notable exceptions being the AL and PC indices and B$_\textrm{Z}$). PCA is a dimensionality reduction algorithm, compressing the feature space by identifying the strongest patterns. The fact that it shows geomagnetic indices as most important is indicative that these indices may contain the information provided by individual variables such as V$_\textrm{sw,x}$. On the other hand, RF is a feature selection algorithm that quantifies importance based on information gained when data are split based on any given feature \cite{Janecek_2005}. The PCA scores, then, quantify a given feature's contribution to the principal components that explain most of the variance in the features, whereas RF quantifies the information gained based on variability of the target variable. Indeed, their difference is why we chose to present their results together and their differences and similarities provide the important collective information for our feature importance approach. Finally, we also acknowledge that the use of non-normalized features may affect the PCA and RF algorithms differently, contributing to differences.

The results in Figure \ref{feature selection results figure} are not definitive, given e.g., that intercorrelations between input features may be important and only recoverable through a detailed information flow or discriminant analysis, and thus the importances revealed must be taken with appropriate reservation. The results do, however, provide one means to assess the potential impact of input parameters that are useful when combined with the existing body of knowledge that is used to build the initial set.


\noindent From our feature importance analysis we discovered that the solar wind velocity features (V$_{\textrm{x}}$/V$_{\textrm{sw}}$) provide redundant information such that only one feature needs to be kept. The data also indicated that AE, being a combination of the AU and AL indices, could also be removed as a redundant input feature. These findings are consistent with intuition and the literature (e.g., \citeA{Borovsky_2020}). These redundant variables are not included in Figure \ref{feature selection results figure}. From the figure, though, we draw several conclusions that are applied in the ML development below: 
\begin{itemize}
    \item The auroral electroject indices (AU/AL) are important, and their time histories appear to provide novel information; 
    \item F$_{10.7}$ is important, likely due to the importance of the preconditioning of the ionosphere in magnetosphere-ionosphere coupling research (e.g., \citeA{Borovsky_2006}), yet our approach does not indicate that information exists in its time history (see the flat curve across time steps in Figure \ref{feature selection results figure}(b)), which is expected because the resolutions examined are beneath the daily temporal resolution of the index. This gives credence to our aggregation process; 
    \item B$_{\textrm{x}}$ and B$_{\textrm{y}}$ are relatively unimportant. We recognize that there are indeed circumstances where B$_{\textrm{x}}$ is important (e.g., \citeA{Laundal_2018}), but those are likely not prominent amidst the other effects in our large database. B$_{\textrm{y}}$ is more generally important (e.g., \citeA{Liou_2018}), but the impact is likely reduced in our need to combine the hemispheres to fill in spatial data coverage;
    \item Newell and Borovsky coupling functions are unimportant when individual solar wind variables and other geomagnetic activity indices are included. This likely points to the fact that this information is contained within those other features, albeit nonlinearly; 
    \item The $Sym$-$H$ index, P$_{\textrm{SW}}$, and B$_{\textrm{z}}$ and their time histories are critical;
    \item There is high variability in feature importances at the $t-$1 h and $t-$45 min time points, indicating that these may contain discriminative information;
    \item B$_{\textrm{z}}$ peaks in importance near $t-$1 h and across time steps has more importance than B$_{\textrm{y}}$ for the analysis that we applied;
    \item The $t-$5 and $t-$3 hour time history data points provide similar information, with some variables decreasing and others increasing across them such that we choose to include only the $t-$3 hour point;
    \item $t-$15 min time history data points do not provide sufficiently novel information (for both PCA and RF the importance is flat from $t-$15 to $t-$5 min, so only one of the points need be kept);
    \item The time points to keep for our modeling are: 6, 3, and one h prior to $t=0$ and 45, 30, and 10 min prior to $t=0$. 
\end{itemize}

After our full feature importance analysis, the final set includes 73 input features. The full and reduced sets of inputs are available at the Github repository for this work. 

\subsection{Final features and machine learning model details} \label{model choice subsection}

\change{All models have parameters that must be chosen in some manner. The ML community refers to these parameters as hyperparameters and their selection, or tuning, determines performance. For instance, the number of layers in the neural network represents one model hyperparameter.}{Like all models, ML models have parameters that must be chosen in some manner, or hyperparameters.} Hyperparameter determination is challenging, requiring both expertise, extensive trial and error, and informed feedback on performance. Exhaustive grid search of the hyperparameter space is inefficient and computationally prohibitive \cite{Bergstra_2012}, but there are methods to create greater efficiency in the search without sacrificing performance based on informed feedback \cite{Bengio_2012, Smith_2018}. The physics-informed and data-driven foundation detailed in the previous section allows our search to be more directed than if taking a purely brute force approach, and we hold some hyperparameters constant noting that for many data sets only a small subset of hyperparameters drive performance \cite{Bergstra_2012}. This paper presents a neural network ML model for particle precipitation. Thus, our search was conducted over a hyperparameter space that included number of epochs (defined as one pass of an entire dataset through the neural network), batch size, number of hidden layers and number of neurons in each hidden layer, the magnitude and placement of dropout layers \cite{Srivastava_2014}, and learning rate. The loss function used is the mean squared error (MSE). 

Assessing the capability of an ML model relies on splitting the data into training and validation sets (sometimes further dividing the training data into training and test sets). We divide our full database into training and validation sets. The training set is used to train the neural network, while the validation set acts as a database not shown to the model at training time and over which to examine model generalizability (i.e., how well the model has learned). \add{Our withheld validation set consists of all DMSP F16 satellite data in the year 2010, more than 50000 data points and over 1500 high-latitude passes.} Taking a consecutive year of data for validation, rather than randomly sampling the full database, prevents influence by autocorrelations in the data (e.g., training with DMSP F16 and using the next data sample as a validation data point). Thus, our choice avoids data `bleeding' that can produce misleading model capability.

The final selected model hyperparameters are: epochs=1000 (with an early stopping criterion to stop training when improvement on the validation set on new epochs ends); batch size=32768; number of hidden layers=8; dropout=0.1 and located after the first hidden layer; and learning rate=0.001. \remove{The network architecture is shown in Figure. The architecture consists of ten layers, eight `hidden' (note that the input layer is not explicitly shown). The number of neurons were tuned and the widths of the green layers in the figure represent them to scale. Our network i}\add{The network architecture consists of ten layers, eight `hidden', and }was designed to create a bottleneck in feature space by first increasing and then decreasing the size of the layers (in terms of number of neurons) \cite{Lamb_2019}, which acts similarly to an autoencoder \cite{Rumelhart_1986} and is useful when input features may exhibit correlations with each other. \add{A visual and further description of the architecture are provided in the Supporting Information file.}

\begin{figure}[h] 
\caption{Observing the model during training. (a) and (b) show the model loss over the training (blue trace) and validation (orange trace) on mean absolute error (MAE) and mean squared error (MSE) as a function of training epochs. (c) shows the histogram of total electron energy flux at four selected epochs overlayed on the actual Defense Meteorological Satellite Program (DMSP) validation observations (black histogram data). The snapshots are for: 1 (red), 25 (orange), 100 (blue), and 360 (green) epochs. Corresponding vertical bars at these epochs are shown on (a) and (b).}
\label{model during training figure}
\centering
\includegraphics[width=0.6\textwidth]{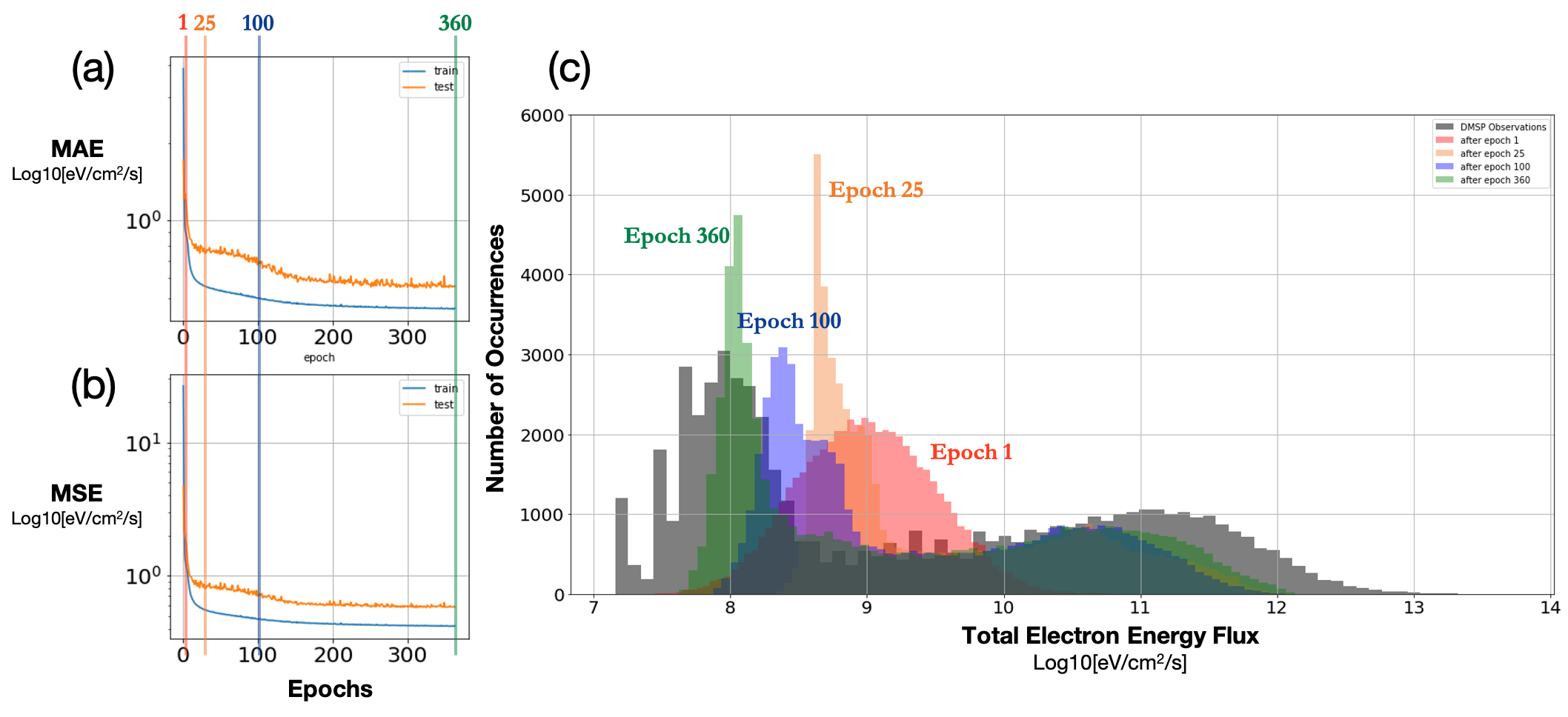}
\end{figure}

Typically model design is guided simply by the loss calculated over the entire validation data set. However, more granular information is needed at training time to inform the design choices. A critical piece of additional feedback is to observe how the model architectures learn during training. We achieve this by observing how well each trained model matched the distribution of the DMSP validation set (chosen to be F16 data collected throughout 2010) across epochs of training. Figure \ref{model during training figure} shows the ability of our final architecture to learn characteristics of the energy flux data during training. Figures \ref{model during training figure}(a) and (b) respectively show the mean absolute error (MAE) and mean squared error (MSE) for the training (blue trace) and validation (orange trace) across training epochs. Epochs refer to the number of times that the learning algorithm works through the entire training dataset. We highlight four epochs, the first, 25th, 100th, and final (360th, after early stopping) epoch as vertical bars on the plots and provide the corresponding histogram of model predictions for each epoch in Figure \ref{model during training figure}(c). In the histograms, we provide the validation data distribution in black, which reveals that the data are clearly bimodally distributed, alluding to the complexity of these data and corresponding to the finding that electron precipitation data are bi- or multi-modally distributed, depending on location and activity level \cite{Hardy_2008}. In Figure \ref{model during training figure}(c) all data are provided in log10$\left[\frac{eV}{cm^2 \cdot s}\right]$ units. The x-axis shows the observational/predicted value while the y-axis provides the number of observations/predictions that fall in that bin. 

After the first pass through the data the model predictions on the validation data show a unimodal distribution centered near the geometric mean of the DMSP validation data at 9.2 log10$\left[\frac{eV}{cm^2 \cdot s}\right]$. This would be like a prediction of all zeros to represent a sine curve and clearly is not skillful. Averages as representations for complex data sets are similarly of limited utility. As the epochs progress, however, the model progressively learns more complex representations of the data and is able to recover the bimodal distribution by epoch 360 (shown in green). The final representation does not capture the diminishing number of samples at the extremes, but the flattening of the training and test MAE and MSE loss curves in Figures \ref{model architecture figure}(a) and (b) indicates that the model is no longer sensitive to new passes through the data and thus has stopped learning from them. The final model, including the trained weights, may indeed be useful to develop an extreme precipitation model through methods like transfer learning \cite[Chapter~9]{Olivas_2009}. 

The corresponding model, which is used for all following results is hereafter labeled \textbf{PrecipNet}.

\section{Comprehensive evaluation and discussion of PrecipNet: Multi-level interrogation and unification of ML and physics understanding (i.e., explainability)} \label{interrogation section}

\citeA{Robinson_2019} thoroughly explored the issue of validation of auroral electrodynamics models. Their findings included: to constrain the number of metrics for auroral electrodynamic parameters, some compromises are inevitable; good metrics are those that reflect overall improvement in model capability for all or most applications; and metrics are most useful when they not only assess the validity and accuracy of models but also provide information about the source of model strengths or weaknesses. They acknowledge the challenge and corresponding absence of `ground truth' data, requiring multi-faceted approaches to evaluation. Thus, we have formulated an interrogation across three levels: 1) standard assessment metrics, a set compiled based on the recommendations above unified with those from \citeA{Liemohn_2018}; 2) comparison with a model deemed the `state of the art' (OVATION Prime for this work); and 3) ability to reproduce known physical phenomena. The framework itself is a key contribution of this work, unifying recommendations for model evaluation across Heliophysics and space weather model evaluation \cite{Bloomfield_2012, Barnes_2016, Liemohn_2018, McGranaghan_2018, Leka_2019}. 



\subsection{Interrogation level \#1: What is the model performance across standard assessment metrics on novel data (i.e., validation data)?} \label{interrogation level one subsection}
We first test the model on standard assessment metrics. The metrics we examine follow \citeA{Liemohn_2018} and are described in Table \ref{metrics table}, consisting of: linear fit intercept and slope, coefficient of determination, root mean square error (RMSE), and mean absolute error (MAE). 

A word of note on coefficient of determination: we define this metric as the Nash-Sutcliffe model efficiency coefficient \cite{Nash_1970}. It is sometimes called prediction efficiency, so we use the label `PE' for this metric. Our definition of PE can be negative, indicating that the mean of the data is a better fit to the observations than the predicted values. Figure \ref{model during training figure}(c) revealed a bimodal nature of the DMSP data in log10$\left[\frac{eV}{cm^2 \cdot s}\right]$ space which is an important characteristic that any model must capture. This definition of PE provides information about model capability to capture such behavior. Further discussion of all metrics can be found in \citeA{Liemohn_2018}.  

    
	\begin{sidewaystable}
	\caption{Model evaluation metrics. In the table, $y$ is the target variable (e.g., the observation), $\hat{y}$ is the model prediction, and $\bar{y}$ is the mean of the observations.}
	\centering
	\begin{tabular}{| l | c | l | l |}
	\hline
	 \textbf{Metric} & \textbf{Equation} & \textbf{Significance/Value}& \textbf{Shortcoming}\\
	\hline
	\multirow{4}{*}{PE} & \multirow{4}{*}{$1-\frac{\sum{(y_i-\hat{y}_i)^2}}{\sum{ (y_i-\bar{y})^2}}$ } & \% of target variable variation & Cannot determine whether \\
	&  &  explained by a &  predictions are biased;\\
	&  &  linear model &  Does not indicate whether\\
	&  &  & model is adequate\\
	\hline
	\multirow{2}{*}{Intercept of linear fit} & \multirow{2}{*}{ $ mx + \textbf{b}$ } & Bias of the model & \\
	&  &   &  \\
	\hline
	\multirow{2}{*}{Slope of linear fit} & \multirow{2}{*}{ $ \textbf{m}x + b$ } & Ability of the model to represent & \\
	&  &  the mean of the observations  &  \\
	\hline
	\multirow{1}{*}{Root Mean Squared } & \multirow{3}{*}{$\sqrt{\frac{1}{N}\sum(y_i-\hat{y}_i)^2}$} & Quadratic mean of the difference  & It is an aggregate and larger  \\
    Error (RMSE) &  & between model and  & errors have a  \\
     &  & observations & disproportionately large effect \\
	\hline
	\multirow{1}{*}{Mean Absolute } & \multirow{2}{*}{$\frac{1}{N}\sum{(|y_i-\hat{y}_i)}|$} & Average absolute different between  & Disregards sign \\
	Error (MAE) & & model and observations & of the error \\
	\hline
	
	\end{tabular}
	\label{metrics table}
	\end{sidewaystable}

Table \ref{model comparison metrics table} summarizes the PrecipNet model over these metrics. For each metric we use the data in units of log10$\left[\frac{eV}{cm^2 \cdot s}\right]$ for the calculation, the same units used to train PrecipNet. The values reported are for the full set of validation data. 

The presence of many interdependent factors in space weather data require models developed with them to provide a measure of uncertainty \cite{Morley_2020}.  Therefore, in Table \ref{model comparison metrics table} and in the time series analyses that follow, we provide this measure for PrecipNet using cross-validation (CV). In CV different models are created from subsets of the data (cross folds) for which each cross fold is a distinct set of data to train and test the model on. For all CV results we choose the cross folds in a stratified manner. That is we create 15 bins for the observations based on the magnitude of the energy flux and ensure that each cross fold set preserves the percentage of samples in each bin. For each cross fold, the full training data are divided such that a unique test set is withheld and the model evaluated over the validation data set. We use ten-fold cross validation (meaning we repeat the process ten times), a general heuristic and deemed robust for Heliophysics applications \cite{Bobra_2015}. The final uncertainty is then the standard deviation across the ten separate realizations of the model. This is like having a panel of `experts' each having developed a different understanding of the problem (developed through different information provided to them) and allowing a better understanding to emerge from the aggregate. The value of this approach has been well described in discussions of ensembling for space weather \cite{Murray_2018, Morley_2018}. 

	\begin{table}[h] 
	\caption{Model evaluation metrics for the machine learning model (PrecipNet) and the `state-of-the-art' OVATION model.}
	\centering
	\begin{tabular}{| l | c | c |}
	\hline
	 \textbf{Metric} & \textbf{PrecipNet} (mean $\pm$ 1-$\sigma$) & \textbf{OVATION}\\
	\hline
	 \textbf{PE} & 0.751 $\pm$ 0.007 & -0.512\\
	\hline
	 \textbf{Intercept of linear fit} &  2.632 $\pm$  0.248 & 7.588\\
	\hline
    \textbf{Slope of linear fit} &  0.717 $\pm$  0.027 & 0.348\\
	\hline
	 \textbf{RMSE} &  0.764 $\pm$  0.011 & 1.887\\
	\hline
    \textbf{MAE} &  0.558 $\pm$  0.019 & 1.574\\
	\hline
	\end{tabular}
	\label{model comparison metrics table}
	\end{table}

Table \ref{model comparison metrics table} reveals that the ML model explains 75\% of the observed variation in the data (PE$=$0.75) and attains an MAE below 0.6. Comparison of these metrics with those of the OVATION Prime model reveals improved performance for the data tested over the metrics examined. The negative PE value for OVATION Prime indicates that the mean of the data provide a better fit to the observations than the model predictions. Figure \ref{histogram results figure}(a) helps further understand this outcome by showing the OVATION Prime predictions of the DMSP validation data as a histogram. We see that OVATION predicts values largely in the high energy portion of the distribution of DMSP validation observations, statistically over-estimating the energy flux, which we discuss below is due to spatially spreading energy flux information more broadly than the observations suggest. Such spreading limits or prevents the capture of mesoscales by many existing particle precipitation models. As a result, we find that a somewhat consistent overestimation of fluxes at sub-auroral and polar cap latitudes is a strong reason for OVATION's negative PE value. 
	
We acknowledge that OVATION uses only solar wind variables as input (the NCF depends on the IMF and the velocity), whereas PrecipNet also include geomagnetic activity indices such that the comparison is not one-to-one. To provide a closer comparison, we created a version of PrecipNet that only uses solar wind input (keeping all time history points). We find that for these summary metrics accuracy is only slightly degraded (the SW-only model attains an MAE of $\sim$0.6). This may seem counter-intuitive or surprising, but further investigation revealed that while the aggregate skill of the model is only moderately degraded, the location of the errors is affected. The SW-only model provides the same improvement at extra-auroral latitudes where fluxes are generally quite low, however within the 60-75$^{\circ}$ MLAT band there is a degradation in model accuracy using only SW inputs. A full comparison is more appropriate for a standalone piece and is the focus of future work.  

\subsection{Interrogation level \#2: What is the model performance against the `state-of-the-art'?} \label{interrogation level two subsection}


 

The summary metrics in Table \ref{model comparison metrics table} can be misleading if not taken with rigorous supporting interrogation. For instance, space weather data are often `quiet' or undisturbed, producing heavily imbalanced data sets \cite{McGranaghan_2018}. A model that predicts nothing but the quiet conditions, missing the dynamics that are most important in space weather modeling, may produce low RMSEs and MAEs, yet may be quite incapable for space weather modeling needs. Thus, all model interrogation must take the important next step to understand the model characteristics beyond summative metrics. 

First, we would like to determine how well the model captures important characteristics in the data, one of which we already described--the bimodal nature. Figure \ref{histogram results figure}(a) shows the histogram (one-dimensional density estimate) of DMSP validation data in black with the same data as determined by the OVATION (red) and PrecipNet (green) models superimposed. Data are shown in log10$\left[\frac{eV}{cm^2 \cdot s}\right]$ units. The x-axis shows the observational/predicted value while the y-axis provides the number of observations/predictions that fall in that bin. Figure \ref{histogram results figure}(b) are two-dimensional Gaussian kernel density estimates for PrecipNet (left) and OVATION (right). The colorbar shows the probability of the model predicting in a given bin across the space of total energy fluxes using a 100x100 bin grid. 

\begin{figure}[h] \label{histogram results figure}
\caption{Density estimates of the model predictions against the DMSP validation data. (a) Histogram of DMSP validation data in blue with the same data as determined by the OVATION Prime (red) and PrecipNet (green) models superimposed. (b) 2D Gaussian kernel density estimate for PrecipNet (left) and OVATION (right). Superimposed are dashed black lines indicating Observation=Prediction. The data in both density estimates are in units of log10$\left[\frac{eV}{cm^2 \cdot s}\right]$ units.}
\label{histogram results figure}
\centering
\includegraphics[width=0.6\textwidth]{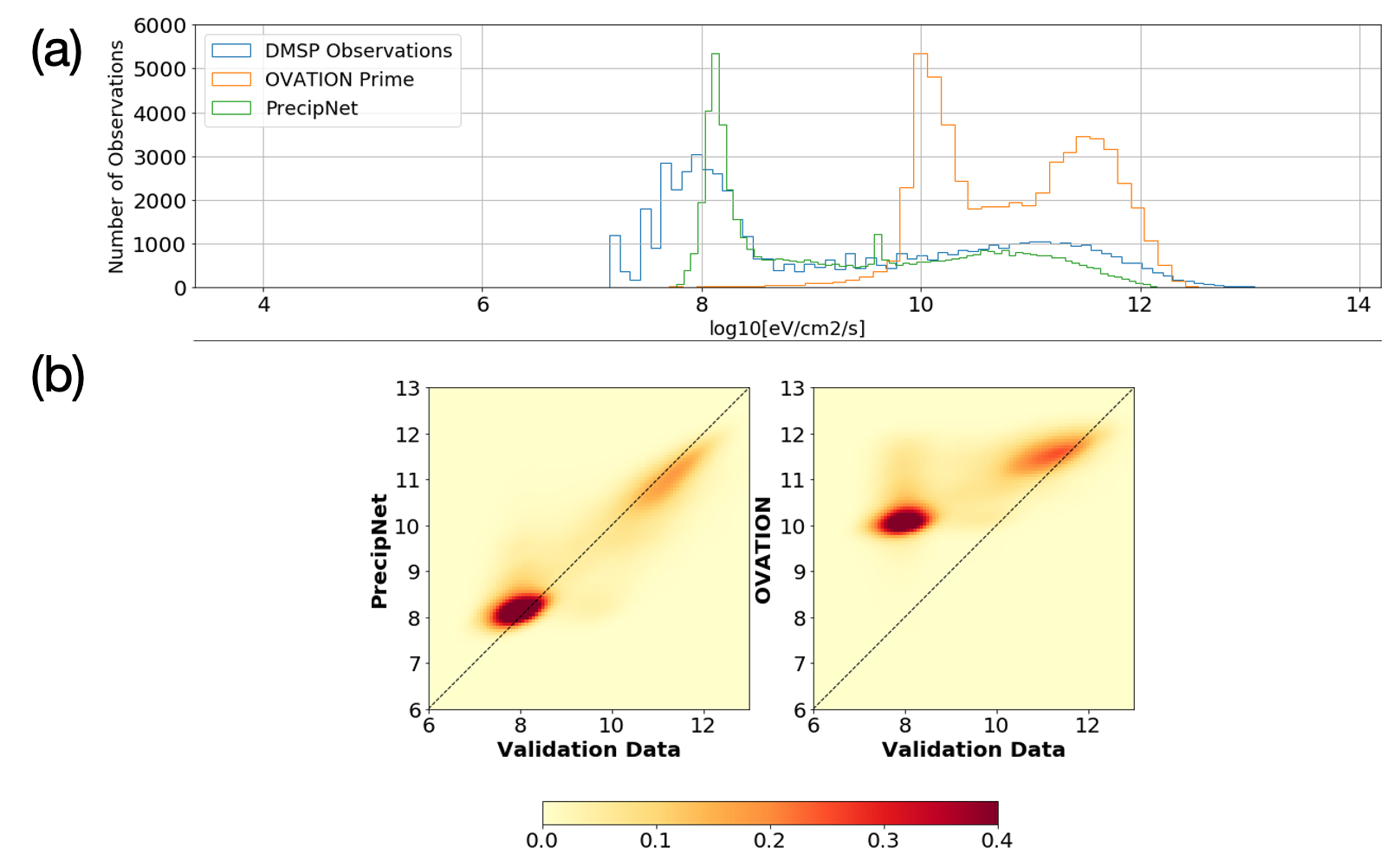}
\end{figure}

It is clear from Figure \ref{histogram results figure}(a) that both OVATION and PrecipNet indicate a bimodal distribution in the data, however only PrecipNet accurately locates the electron energy flux values (x-axis) of the peaks in the distribution of the DMSP validation data that are provided in blue. OVATION's overestimation of the energy flux is indicative of a spreading of information that results in an inability to capture dynamic boundaries that emerge between low and high precipitation. PrecipNet distinguishes the low and high flux regions by their appropriate magnitudes. 

The improved agreement is supported by the two-dimensional density estimate in Figure \ref{histogram results figure}(b). Observation=Prediction lines are superimposed on the color maps (dashed black lines). PrecipNet and OVATION have Pearson's correlation coefficients of 0.87 and 0.70, respectively.

\begin{figure}[h] 
\caption{Comparing PrecipNet and OVATION Prime for the validation data time series. For all data points in the validation set (DMSP F16 throughout 2010) PrecipNet and OVATION Prime are evaluated at the location of the satellite. We select a representative time period (November 14, 2010 17:00--21:00 UT). (a) Three-panel plot for the full time range with the top panel showing the DMSP observations (blue trace) and OVATION output (red trace), the middle panel plotting the same DMSP observations with PrecipNet predictions (median as the solid line and one-sigma standard deviation as the envelope from the cross-validation results), and the bottom panel showing the residuals for both models (PrecipNet as green points and OVATION as red); (b) Three-panel plot showing the same information, except zoomed into 19:30---20:00 UT); and (c) the polar ground track of DMSP F16 during the zoomed time.}
\label{time series figure}
\centering
\includegraphics[width=0.6\textwidth]{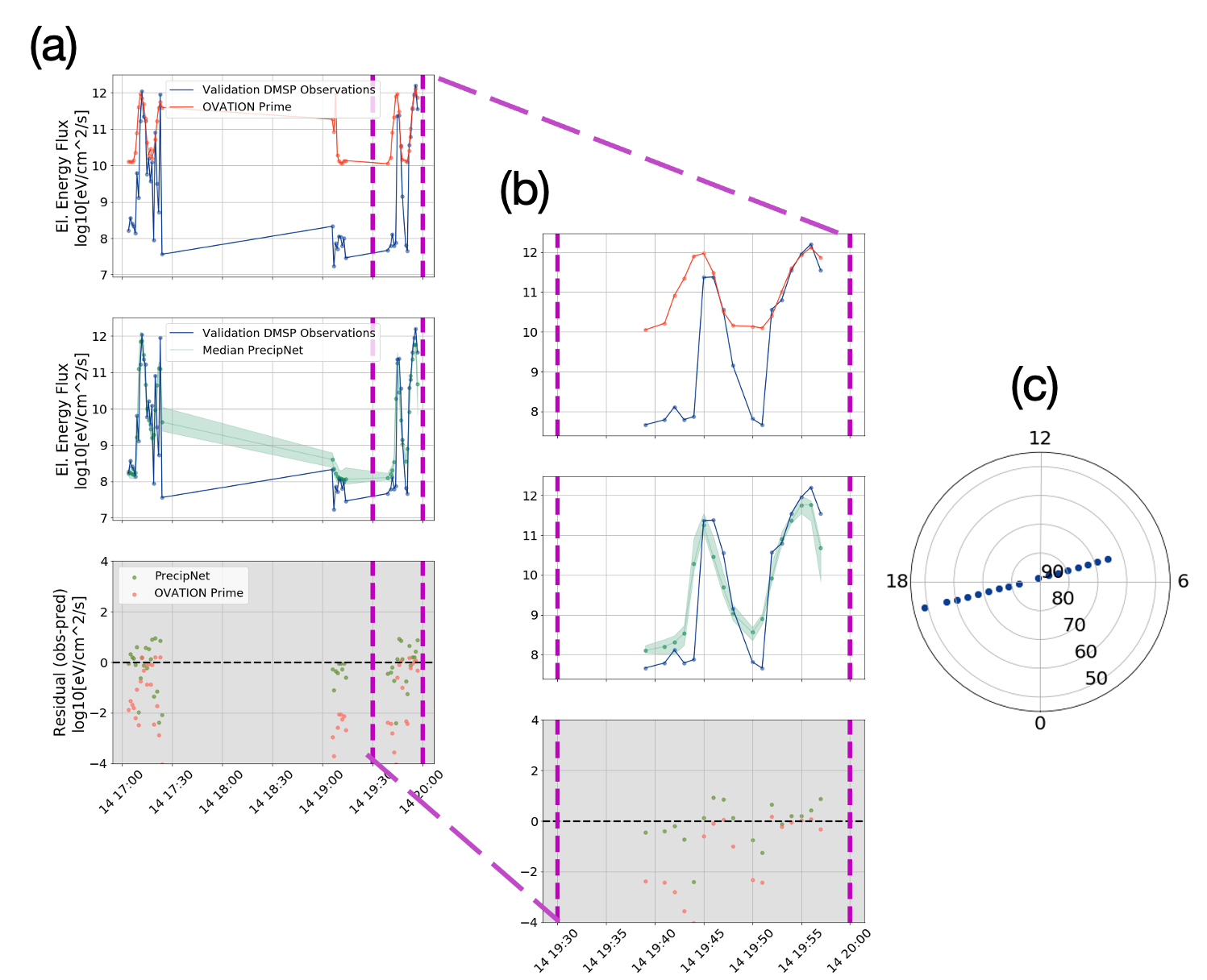}
\end{figure}

The trends from studying the density estimates of validation data \add{(which represent more than 1500 DMSP F16 passes}) can be more clearly and granularly seen in Figure \ref{time series figure}, which shows the PrecipNet and OVATION Prime model results as compared with the validation data on time series. \add{For this paper, w}e select \change{a}{one} representative time period (November 14, 2010 17:00-21:00 UT) and provide results over the full time range (Figure \ref{time series figure}(a)) where the top panel illustrates the DMSP observations (blue trace) and OVATION output (red trace), the middle panel plots the same DMSP observations with PrecipNet predictions (median as the solid line and one-sigma standard deviation as the envelope from the CV results), and the bottom panel shows the residuals for both models (PrecipNet as green points and OVATION as red). The tendency toward overestimation by OVATION is clearly demonstrated where low flux values are biased upward and corresponding dynamic boundaries are inaccurately specified. PrecipNet more accurately specifies the low fluxes and the mesoscale transition to high fluxes throughout the time period (notice the rise at $\sim$0943 UT takes about one minute for PrecipNet ($\sim$400 km) but roughly 3-4 minutes for OVATION). The residuals (observation$-$prediction) on the bottom panel for PrecipNet (green points) are scattered around zero with little structure whereas there is both bias and structure in the OVATION residuals (blue points). Figure \ref{time series figure}(b) zooms in on a single pass from 19:30-20:00 UT, the ground track for which is shown in the inset Figure \ref{time series figure}(c). The single pass more closely shows the accuracy of PrecipNet's reconstruction of the energy flux and uncertainty bounds that nearly contain the extrema in the data. \add{Further case studies, including multi-platform observations, will be the focus of follow-on work.}

\subsection{Interrogation level \#3: Can the model reproduce known physical phenomena?} \label{interrogation level four subsection} 


One of the most important aspects of the ML approach is the ability to create global, time-dependent specification and prediction \cite{Bortnik_2016} to evaluate the model performance on known physical phenomena. Figure \ref{january 25, 2000 event figure} reveals an analysis of the PrecipNet model to reconstruct global high-latitude maps of total electron energy flux on January 25, 2000, an event selected for the geomagnetic and substorm conditions and the data coverage by the Global Geospace Science Polar satellite Visible Imaging System (VIS) \cite{Frank_1995}. 

\begin{figure}[h] 
\caption{Model interrogation of known physical phenomena. Results of the PrecipNet and OVATION models are produced for seven discrete times throughout January 25, 2000. The top time series panel shows the interplanetary magnetic field (IMF) x- (green), y- (blue), and z-components (red). The next time series shows the auroral electroject (AE) index.  During steps D--G a substorm occurs, beginning at 17:45 UT at 68.71$^{\circ}$ magnetic latitude (MLAT) and 23.13 hours magnetic local time (MLT) as observed by the SuperMagnetometer initiative substorm database. At all steps indicated by vertical purple lines on the time series we provide global polar high-latitude maps for PrecipNet (left) and OVATION (right). During the substorm, Polar satellite Visible Imaging System (VIS) data are available and are included (far right). All polar plots show the northern hemisphere with the sun to the top of the figure, MLT around the dial, and MLAT extending to 50$^{\circ}$ and white rings in increments of 10$^{\circ}$. Superimposed on the model polar plots are the auroral region boundaries (polar=bright blue and equatorward=bright yellow) as determined by the 0.2 $\frac{erg}{cm^s\cdot s}$ threshold. The hemispheric power estimates for each model map are provided at the top of the polar plot. The color bar at the bottom shows the color scale for all PrecipNet and OVATION maps (log10$\left[\frac{eV}{cm^2 \cdot s}\right]$), while the Polar VIS data are shown in units of Rayleighs.}
\label{january 25, 2000 event figure}
\centering
\includegraphics[width=0.6\textwidth]{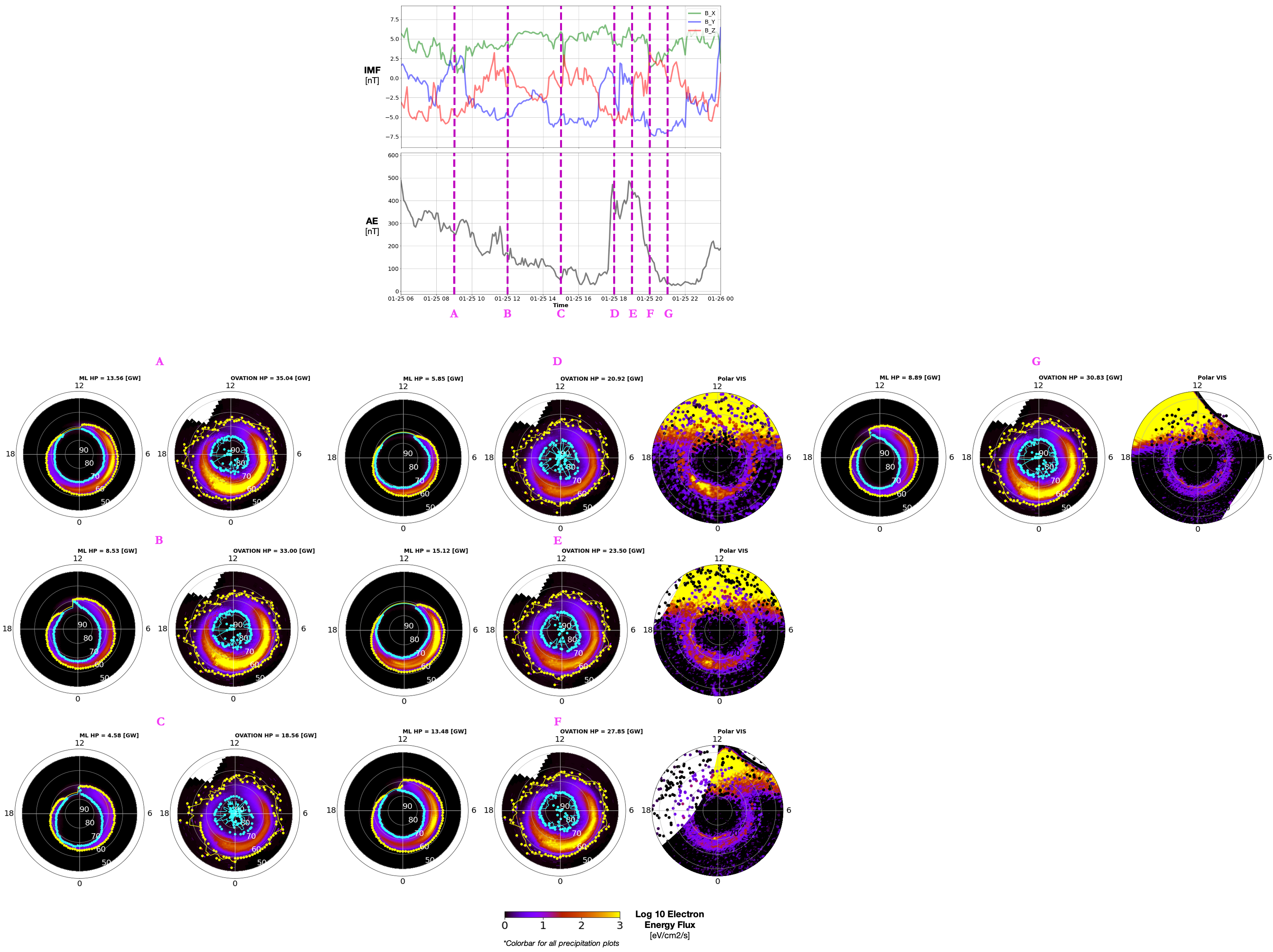}
\end{figure}

Jan 25, 2000 is a unique period due to the range of conditions and phenomena it covers, including gradual changes geomagnetic activity and an isolated substorm. Coupled with the availability of quality Polar VIS observations during the substorm, it is an ideal time period for evaluation. For seven selected snapshots (A--G) indicated by vertical purple lines on the time series plots in Figure \ref{january 25, 2000 event figure}, we provide global polar high-latitude maps for PrecipNet (left) and OVATION (right). The top time series panel shows the x- (green), y- (blue), and z-components (red) of the IMF and the second from the top is the time series of the AE index. All polar plots show the northern hemisphere with the sun to the top of the figure, MLT around the dial, and white rings in 10$^{\circ}$ MLAT increments down to 50$^{\circ}$. Superimposed on the model polar plots are the auroral region boundaries (polar=cyan and equatorward=bright yellow) as determined by a 0.2 $\frac{erg}{cm^2 \cdot s}$ threshold independently for each MLT bin at each UT. This is an approach identified by NASA's Community Coordinate Modeling Center's auroral model evaluation plan and is related to findings from \cite{Zhang_2008}. If the 0.2 threshold is not exceeded, no boundary is indicated. In the title to each polar plot we provide estimates of the total electron precipitation hemispheric power for each UT, obtained by integrating over 50--90 MLAT and over all MLTs. Note that there are several approaches to making this calculation (particularly with how to calculate the hemispherically integrated area), with little to no ground-truth hemispheric power to compare against, so we avoid absolute evaluation of the HP values, focusing instead on a relative comparison between PrecipNet and OVATION Prime. Finally, both PrecipNet and OVATION Prime are evaluated and visualized using the native OVATION grid (0.5$^{\circ}$ MLAT $\times$ 0.25 MLT) for consistency. PrecipNet can be evaluated at arbitrary resolution. 
 
Over 06:00--15:00 UT there is a gradual decrease from a geomagnetically active condition to extreme quiet, as indicated by the steadily decreasing AE index and the turning of the IMF z-component to near zero or northward conditions. Polar plots for snapshots A--C show that both models track the contracting auroral oval both in intensity and location of the oval. PrecipNet's equatorward boundary moves from below 60$^{\circ}$ MLAT at midnight MLT to near 65$^{\circ}$. As is characteristic over all times, OVATION specifies a broader auroral oval with both higher MLT-to-MLT bin variability at a given UT snapshot and smaller variation across UT snapshots (cf. the nightside equatorward oval remains relatively stationary over the snapshots). This behavior reveals two important points: 1) PrecipNet captures more dynamic auroral boundary behavior that is important to dynamics of the high-latitude auroral region; and 2) there are periods during which PrecipNet's specification of the breadth of the oval may be too slight, resulting in smaller estimates of the hemispheric power. For both models, moving into extreme quiet from B to C there is roughly a 50\% reduction in the hemispheric power estimates.   

During steps D--G, a substorm occurs, beginning at 17:45 UT and 68.71$^{\circ}$ magnetic latitude (MLAT) and 23.13 hours magnetic local time (MLT) as observed by the SuperMagnetometer initiative substorm database \cite{Gjerloev_2011}. For each of these steps we also display the available Polar VIS global auroral emission data [kRayleighs] which observed this substorm. At step D, Polar VIS data show a brightening centered at 65$^{\circ}$ MLAT and 22 MLT that is not prominently seen in either auroral model. However, PrecipNet indicates a faint increase across the premidnight to 06 MLT sectors at 65$^{\circ}$ MLAT roughly corresponding to the VIS data. There is no discernible increase in the OVATION output. At step E, the AE index and VIS data indicate that the substorm is sustained. Here PrecipNet specifies a large increase in the breadth of the auroral oval, expanding both equatorward and poleward consistent with known oval dynamics during activity \cite{akasofu_2017} that is not captured by OVATION Prime, and a brightening in the premidnight MLT sector with striking correspondence to Polar VIS. From E to F the OVATION auroral oval grows in intensity while PrecipNet and Polar VIS indicate a waning corresponding to the significantly reduced AE index. The snapshots taken throughout the substorm provide evidence that PrecipNet is capable of mesoscale and particularly substorm-scale specification.

It is important to note that we do not expect a one-to-one correspondence between the particle precipitation maps and the Polar VIS imagery maps. Rather the correspondence is qualitative and provides insight specifically to important features and auroral behavior. This is due to a number of effects, including DMSP capturing precipitation that does not make visible light, convolved effects in the auroral imagery data, and different temporal and spatial resolutions between the imagery and the model. For instance, the Polar VIS data are sensitive to emissions from both electron and proton precipitation \cite{Frank_1995}, though PrecipNet only quantifies electron precipitation. For instance, in Figure \ref{january 25, 2000 event figure}E PrecipNet specifies an enhancement in the 04-06 MLT sector that is not as pronounced in the Polar VIS imagery. Precipitation in this sector is predominately diffuse, which may drive less intense emissions. We suspect that the cause of such differences is a combination of the reasons listed above and will be investigated further with detailed case studies that incorporate new platforms such as incoherent scatter radar to more fully diagnose. Nonetheless, to zeroth order, Polar VIS brightnesses are proportional to energy flux. The PrecipNet model more faithfully represents fine details apparent in the imagery data, including a brightening across midnight sector local times at 1900 UT when Polar VIS observes strong activity associated with a classic substorm in that sector (Figure \ref{january 25, 2000 event figure}E). 



By evaluating PrecipNet in the context of physical phenomena, we allow the performance to be judged through physical understanding. This represents a layer of explaining the ML results, and allows physics-informed improvement of the model. This is one method to accomplish ’theory-guided data science’ \cite{Karpatne_2016} and ‘computer-aided discovery’ \cite{Pankratius_2016}, critical areas of development for the ’New Frontier’ of scientific discovery in the digital age \cite{McGranaghan_2017}. 

\subsection{Discussion: Identifying and facilitating next steps for a community of particle precipitation modeling} \label{usability section}

We have produced a new model of electron precipitation making new use of cutting-edge tools and capabilities from the fields of data science and ML. We are quite aware that previous efforts have produced similarly profound advances. These efforts include, for instance, those of three other working groups across the community: 1) The Auroral Precipitation and High Latitude Electrodynamics (AuroraPHILE) Community Coordinated Modeling Center (CCMC) working group; 2) the G1-03 (Auroral Electrodynamics) Working Team under the G1 Cluster (Geomagnetic Environment) as part of the Committee on Space Research (COSPAR) International Space Weather Action Team (ISWAT); and 3) the Geospace Environment Modeling (GEM)-Coupling, Energetics, and Dynamics of Atmospheric Regions (CEDAR) Conductance Challenge Working Group. 

Our approaches already identify areas of future exploration, including other means of assessing the importance of features (Section \ref{feature importance subsection}) and examining ML model hyperparameters (Section \ref{model choice subsection}). However, to align this work with the ongoing community efforts we also provide suggestions for particularly important areas of next investigation: 
\begin{itemize}
    \item Introduce new precipitation databases: much like the series of OVATION models, it is clear that improvement is possible through an increase in coverage of input data (both spatially and in terms of geomagnetic activity). Our approach is naturally extensible to any other data that provide energy flux as a function of position. The only task is to prepare those data in the same format as the input samples described above (see Figure \ref{data preparation figure}). An immediate next task is to grow the model with the Fast Auroral SnapshoT (FAST) Explorer electron precipitation database, accounting for differences arising from the distinct orbits of the FAST and DMSP spacecraft. 
    
    \item Extend the model using magnetometer data: There is strong evidence that local measurements from magnetometers provide information about local precipitation behavior \cite{Newell_2011, Newell_2015}. Therefore, an important extension of this work is the incorporation of magnetometer data. The structure of our data preparation and analyses is indeed particularly designed to facilitate such extensions. One only has to determine the local magnetometer information, either through conjunctions with DMSP observations and ground-based magnetometer stations or through the use of several `beacon' magnetometer measurements (a subset of magnetometers chosen to be representative of different MLAT-MLT sectors). Using local measurements may recover information that is lost in the aggregation of those data into global indices (e.g., Kp). 
    
    \item Expand ML model exploration: Interaction with the ML community through programs such as the Frontier Development Laboratory (\url{https://frontierdevelopmentlab.org/}) reveals a wealth of potentially valuable ML approaches for the particle precipitation challenge. Three have emerged as particularly high potential from our investigation: 1) treating physical models as first class data sources and using their output as an input feature for an ML model (a `gray-box' approach \cite{Camporeale_2020}); 2) developing probabilistic ML models (e.g., \cite{Camporeale_2019}); and 3) encoding physical constraints into the loss function of the ML model as a means to allow physics to inform the process. 
\end{itemize}



\section{Conclusion} \label{conclusion section}
We have produced a new nowcast model of total electron energy flux precipitation using 51 satellite years of Defense Meteorological Satellite Program (DMSP) particle precipitation data and a set of $>$150 organizing parameters from solar wind and geomagnetic activity indices and their time histories (of which we determine a smaller set that retains the important information to specify the energy flux). Particle precipitation data contain the complexity of the solar wind-magnetosphere-ionosphere system, which dictates the development of models with higher expressive capacity. Further, previous analysis of precipitation models have revealed that the most important aspect of particle precipitation models is the choice and representation of the organizing parameters \cite{Newell_2015}. Thus, the physical situation dictates the use of machine learning algorithms that can respond to both needs. We have explored these algorithms and produced a new model, PrecipNet, that reduces errors over unseen validation data by 50\% from the existing state-of-the-art, deemed here to be the model used in operations, OVATION Prime. \add{We note that there are challenges to using the DMSP data for global models, such as small mid-latitude gaps in magnetic local time coverage post-midnight and post-noon. We have provided a process for creating `analysis-ready data' for particle precipitation ML that will allow other datasets to be integrated into this work, helping to address observational issues.}

We proposed and demonstrated a new model interrogation framework across three levels. In each, PrecipNet provides improvement. First in more accurately representing the complexity inherent in the precipitation data indicated by their bimodal nature \cite{Hardy_2008} (Figure \ref{histogram results figure}), then in recovering sharper dynamic changes in the flux time series data (Figure \ref{time series figure}), and finally across comparisons with auroral imagery during a known substorm event (Figure \ref{january 25, 2000 event figure}). 

We commented on numerous places in the design of ML models where physics can inform or be embedded, providing a guide for future work that dictates the use of ML in Heliophysics. Our results provide a model that can be used to study auroral phenomena, drive high-latitude auroral electrodynamics models, and be used as an input to global circulation models.

\acknowledgments
We provide software and meta-information needed to reproduce analyses in this manuscript and to extend the research at \url{https://github.com/rmcgranaghan/precipNet}. That repository contains information on data preparation, software dependencies needed, the final precipNet model used for this manuscript, and reference to the published data set created and used by this research.

We gratefully acknowledge the International Space Sciences Institute (ISSI) in Bern, Switzerland (\url{https://www.issibern.ch}) who funded the proposal ``Novel approaches to multiscale geospace particle transfer: Improved understanding and prediction through uncertainty quantification and machine learning,'' which originally brought this group together, provided resources for two in-person research weeks, and is a regular source of inspiration for cutting-edge space science research and collaboration. We also acknowledge two members of the first week of those meetings for their knowledge and direct contributions to this work; Irina Zhelavskaya and Yuri Shprits.

Work at the Birkeland Center for Space Science is funded by the Research Council of Norway/CoE under contract 223252/F50 and by ESA contract 4000126731 in the framework of EO Science for Society.

We acknowledge the University of Colorado Boulder Space Weather Technology, Education, and Research Center (SpWx TREC) for providing computational resources that in-part aided our machine learning exploration. 

R. McGranaghan and J. Ziegler were in-part supported under the Defense Advanced Research Projects Agency (DARPA) Space Environment Exploitation (SEE) grant to the Heliosphere to Earth's Atmosphere Rendering via Excellent Artificial Intelligence Training (HEARTBEAT) team under Department of the Interior Award D19AC00009 to Georgia Institute of Technology with subaward to ASTRA. 

E. Camporeale is partially funded by NASA under grant 80NSSC20K1580.

T. Bloch is funded by Science and Technology Facilities Council (STFC) training grant number ST/R505031/1.

We gratefully acknowledge the OMNI initiative, from which we used data from the full set: ``OMNI High resolution (1-min, 5-min) OMNI: Solar wind magnetic field and plasma data at Earth's Bow Shock Nose (BSN)'' and are freely an openly available at \url{https://omniweb.gsfc.nasa.gov/ow_min.html}.

We are grateful to those involved in preparing and providing the Defense Meteorological Satellite Program (DMSP) data over the several decade lifetime of these spacecraft. The data were obtained from NASA's CDAWeb system and are openly and freely available from \url{https://cdaweb.gsfc.nasa.gov/pub/data/dmsp/}. 

We thank the SILSO service at the Royal Observatory of Belgium in Brussels for providing the monthly mean sunspot numbers. Those data can be obtained from \url{http://www.sidc.be/silso/home}. 

For the ground magnetometer data we gratefully acknowledge: INTERMAGNET, Alan Thomson; CARISMA, PI Ian Mann; CANMOS, Geomagnetism Unit of the Geological Survey of Canada; The S-RAMP Database, PI K. Yumoto and Dr. K. Shiokawa; The SPIDR database; AARI, PI Oleg Troshichev; The MACCS program, PI M. Engebretson; GIMA; MEASURE, UCLA IGPP and Florida Institute of Technology; SAMBA, PI Eftyhia Zesta; 210 Chain, PI K. Yumoto; SAMNET, PI Farideh Honary; IMAGE, PI Liisa Juusola; Finnish Meteorological Institute, PI Liisa Juusola; Sodankylä Geophysical Observatory, PI Tero Raita; UiT the Arctic University of Norway, Tromsø Geophysical Observatory, PI Magnar G. Johnsen; GFZ German Research Centre For Geosciences, PI Jürgen Matzka; Institute of Geophysics, Polish Academy of Sciences, PI Anne Neska and Jan Reda; Polar Geophysical Institute, PI Alexander Yahnin and Yarolav Sakharov; Geological Survey of Sweden, PI Gerhard Schwarz; Swedish Institute of Space Physics, PI Masatoshi Yamauchi; AUTUMN, PI Martin Connors; DTU Space, Thom Edwards and PI Anna Willer; South Pole and McMurdo Magnetometer, PI's Louis J. Lanzarotti and Alan T. Weatherwax; ICESTAR; RAPIDMAG; British Artarctic Survey; McMac, PI Dr. Peter Chi; BGS, PI Dr. Susan Macmillan; Pushkov Institute of Terrestrial Magnetism, Ionosphere and Radio Wave Propagation (IZMIRAN); MFGI, PI B. Heilig; Institute of Geophysics, Polish Academy of Sciences, PI Anne Neska and Jan Reda; University of L’Aquila, PI M. Vellante; BCMT, V. Lesur and A. Chambodut; Data obtained in cooperation with Geoscience Australia, PI Andrew Lewis; AALPIP, co-PIs Bob Clauer and Michael Hartinger; SuperMAG, PI Jesper W. Gjerloev; Data obtained in cooperation with the Australian Bureau of Meteorology, PI Richard Marshall.

For the Polar VIS Earth Camera images we gratefully acknowledge the Polar VIS Earth Camera team, PI Prof. Louis A Frank. We further acknowledge and thank from Robin Barnes and the SuperMAG team for providing the VIS image data. SuperMAG includes the entire catalog of images obtained by the Polar VIS instrument. The images and the necessary software was generously made available to the SuperMAG team by the VIS instrumentation team.


%
%






\end{document}